\documentclass[preprint]{aastex}

\bibpunct[]{(}{)}{;}{a}{}{,}

\newcommand{\etal}{et al.\ }
\newcommand{\kms}{\, {\rm km\, s}^{-1}}
\newcommand{\mpc}{\, {\rm Mpc}}
\newcommand{\lya}{Ly$\alpha$ }

\newcommand{\gmo}{\gamma-1}
\newcommand{\bF}{\bar{F}}
\newcommand{\Db}{\Delta_g}
\newcommand{\TDr}{$T-\Db$ relation }
\newcommand{\hi}{\mbox{H\,{\scriptsize I}\ }}
\newcommand{\heii}{\mbox{He\,{\scriptsize II}\ }}
\newcommand{\mmr}{Paper I~}

\shorttitle{Temperature-Density Relation in the IGM}
\shortauthors{McDonald \etal}

\begin{document}

\title{A Measurement of the Temperature-Density Relation in the 
Intergalactic Medium
Using a New \lya Absorption Line Fitting Method
\altaffilmark{1}}

\author{Patrick McDonald,\altaffilmark{2,3}
Jordi Miralda-Escud\'e,\altaffilmark{3,4}
Michael Rauch,\altaffilmark{5} 
Wallace L. W. Sargent,\altaffilmark{6} 
Tom A. Barlow,\altaffilmark{6}
and Renyue Cen\altaffilmark{7}
}

\altaffiltext{1}{The observations were made at the W.M. Keck 
Observatory, which is 
operated as a scientific partnership between the California Institute 
of 
Technology and the University of California; it was made possible by
the generous support of the W.M. Keck Foundation. }
\altaffiltext{2}{Department of Physics and Astronomy, University of 
Pennsylvania, Philadelphia, PA 19104}
\altaffiltext{3}{Department of Astronomy, The Ohio State University, 
Columbus, OH 43210; mcdonald,jordi@astronomy.ohio-state.edu}
\altaffiltext{4}{Alfred P. Sloan Fellow}
\altaffiltext{5}{ESO, Karl-Schwarzschild-Str. 2, 85748 Garching, 
Germany}
\altaffiltext{6}{Astronomy Department, California Institute of 
Technology, Pasadena, CA 91125}
\altaffiltext{7}{Princeton University Observatory, Peyton Hall, 
Princeton, NJ 08544}

\begin{abstract}
The evolution of the temperature in the intergalactic medium is related to the
reionization of hydrogen and helium, and has important consequences for our
understanding of the \lya forest and of galaxy formation in gravitational
models of large-scale structure. We measure the temperature-density relation of
intergalactic gas from \lya forest observations of eight quasar spectra
with high resolution and signal-to-noise ratio, using a new line fitting
technique to obtain a lower cutoff of the distribution of line widths from
which the temperature is derived. We carefully test the accuracy of this
technique to recover the gas temperature with a hydrodynamic simulation. The
temperature at redshift $\bar{z}=$(3.9, 3.0, 2.4) is best determined at
densities slightly above the mean: $T_\star=$($20200\pm 2700$, $20200\pm 1300$,
$22600\pm 1900$)K (statistical error bars) for gas density (in units of the
mean density) $\Delta_\star=$($1.42\pm 0.08$, $1.37\pm 0.11$, $1.66\pm 0.11$).
The power-law index of the temperature-density relation, defined by $T=T_\star
(\Db/\Delta_\star)^{\gmo}$, is $\gmo=$($0.43\pm 0.45$, $0.29\pm 0.30$, 
$0.52\pm 0.14$) for the same three redshifts. The temperature at the fixed
over-density $\Delta=1.4$ is $T_{1.4}$=($20100\pm 2800$, $20300\pm 1400$,
$20700\pm 1900$)K. These temperatures are higher than expected for
photoionized gas in ionization equilibrium with a cosmic background,
and can be explained by a gradual additional heating due to on-going
\heii reionization.
The measurement of the temperature reduces one
source of uncertainty in the lower limit to
the baryon density implied by the observed mean flux decrement.
We find that the temperature cannot be reliably
measured for under-dense gas, because the velocities due to expansion always
dominate the widths of the corresponding weak lines.

\end{abstract}


\section{INTRODUCTION}

The Lyman-$\alpha$ forest absorption in the spectra of quasars provides
a wealth of information about the properties of the intergalactic 
medium (hereafter, IGM).
There has recently been a lot of interest in using the distribution of
Doppler parameters of fitted absorption lines, measuring the total
velocity dispersion of the gas, to constrain 
the temperature of the IGM \citep*{stl99,rgs99,bm99}. 
In this paper we develop a method to identify and to fit absorption
lines, and to obtain the gas temperature from the distribution of the
Doppler parameters of the lines. Our algorithm is intended to be 
simple to implement and be applied identically on simulations and 
observations.

The temperature of the IGM as a function of 
density is primarily determined by the balance between adiabatic 
cooling
and photoionization heating, once ionization equilibrium with the
background radiation has been established. However, during the epoch of
reionization, the heating rate is higher because every atom needs to
be ionized once (and the ionization can occur on a short time-scale
compared to the recombination rate), and the high opacity of the
low-density IGM implies that high-frequency photons are absorbed,
delivering a much greater amount of heat for each ionization
\citep[e.g.,][]{mr94,hg97,hs98,ah99,g99}. Other sources of heating
may also contribute,
such as Compton heating by the X-ray background \citep{me99}, or 
photoelectric heating by dust grains \citep*{nss99}. 
Constraining these sources of heating is one of the two primary reasons
why we are interested in measuring the temperature. The other reason
is the need to make accurate predictions for the statistics of the
\lya forest flux in order to constrain cosmological parameters 
\citep*[e.g.,][]{rms97,wmh97,cwk98,h99,mm99,hsb99,cwp99,wch99,chd99,
nh99,mmr99}.
The temperature-density relation affects the predicted relationship
between the power spectrum of the transmitted flux and the power
spectrum of the initial mass density perturbations 
\citep[]{nh99}, as well as 
the predicted mean 
transmitted flux \citep{rms97,mmr99}, which can be used to constrain
the baryon density of the universe. 

  Recent \lya forest simulations have shown that, when the structure 
of the absorption systems is adequately resolved,
the predicted absorption line widths are smaller than observed
if the temperature of the IGM is determined from
photoionization equilibrium, well after reionization has ended
\citep{tle98,tls99,bma99}. To solve the discrepancy the temperature 
apparently needs to be higher. Several authors have presented
measurements of the IGM temperature using different methods, generally
finding values moderately higher than expected from photoionization
equilibrium \citep{tsh99,rgs99,bm99,str00}.

  Our aim in this paper is to provide a new unambiguous measurement
of the temperature, making a more exhaustive analysis than in previous
work of the model uncertainties that result from comparing the
observational results with a simulation. We develop a new
line-fitting method as an alternative to the standard Voigt-profile
fitting with line deblending, which is much faster, unambiguous, and
easy to implement. Our method works by essentially assigning one line
to each sufficiently deep minimum in the transmitted flux, and
measuring the line width and central optical depth for each line.
The gas temperature at each density is then derived from the distribution
of line widths at each central optical depth.
The systematic uncertainties and model dependence of the method used to
derive the temperature are carefully analyzed, in a more extensive way than
it was done in previous work. The new method is applied to
observational data and to a simulation in exactly the same way,
computing error bars due to the variance in our observed sample.

  The main idea of the method to measure the temperature of the IGM was
suggested by \citet{stl99}, \citet{rgs99}, and \citet{bm99}. The
probability distribution of Doppler parameters, $P(B)$, is characterized
by a lower cutoff, $B_C$, where $P(B)$ rises sharply, with very few
lines having narrower Doppler parameters than this cutoff. The idea is
that this cutoff is a measure of the gas temperature. In general,
absorption lines have both a thermal and a hydrodynamical contribution
to their breadth; however, for any set of lines with similar gas
temperature, the narrowest ones will be those where the velocity field
along the line of sight through the absorber is close to a caustic, so
that the variation in the fluid velocity is minimized and thermal
broadening dominates the observed line width. In fact, it was found by
\cite{tsh99} that the narrowest absorption lines are primarily thermally
broadened.

  A tight relationship between density and temperature in the IGM for
gas at low densities, where shock heating is not very important, is
expected theoretically and is found in numerical simulations
\citep{hg97,tle98}. This implies that the narrowest lines at a given gas
density (corresponding approximately to the optical depth at the line
center) are not selected to have low gas temperature, but low fluid
velocity dispersion. This justifies estimating the temperature from the
lower cutoff of the Doppler parameter distribution.

In \S 2 we briefly describe the observational data and the simulation
that we use.
In \S 3 we describe our line fitting algorithm.
In \S 4 we demonstrate how the line fitter works by running it on 
spectra from the numerical simulation.  
In \S 5 we describe our method for estimating the temperature from
Doppler parameter distribution, testing the conditions under which the
temperature can be recovered in a model-independent way.
In \S 6 we use the line fitter on the observational data 
and give results for the measured temperatures.  
The results are discussed in \S 7.
The Appendix describes further details of our line-fitting method.

\section{THE OBSERVATIONAL DATA AND THE SIMULATION}

\subsection{Observations}

We use the same set of eight quasar spectra as in \mmr.
These spectra have sufficiently high resolution and signal-to-noise 
ratio to measure the shape of each absorption feature.
The pixel noise is typically less than 5\% of the continuum flux level,
and frequently as low as 1\%.
The velocity resolution is 6.6 $\kms$ (FWHM) and the spectra are 
binned in 0.04 \AA~ pixels.  More details and statistics of this data 
set are given in \mmr and references therein.

The seven quasars from the \citet{rms97} data set have previously
constructed lists of
regions that are suspected of containing metal lines.  Our main 
results include these regions in the spectra
because they are not positively 
identified as containing metal lines.

In \mmr we defined three redshift bins:  $3.39<z<4.43$,
$2.67<z<3.39$, and $2.09<z<2.67$, with mean redshifts $\bar{z}=3.9$,
$\bar{z}=3.0$, and $\bar{z}=2.4$.  We use these same three bins in 
this paper, which contain approximately the same amount of data.   

\subsection{Simulation}

We test the profile fitting code and the
procedure to measure the temperature
on the output of the Eulerian hydrodynamical simulation described in 
\citet{mco96} (referred to as L10 in that paper).
The cosmological model used has
$\Omega_0=0.4$, $\Omega_\Lambda=0.6$, $h=0.65$, $\sigma_8=0.79$, and
large scale primordial power spectrum slope $n=0.95$.  The box size 
of the
simulation is $10 h^{-1}$ Mpc, and it contains $288^3$ cells.
We use outputs from the simulation at $z=$4, 3, and 2.

\subsubsection{Generation of Simulated Spectra}

\lya spectra are computed for a large number of lines of sight along
the box axes.
There is one free parameter that we can vary when computing the 
spectra,
the normalization of the optical depth, which we adjust to reproduce 
the
mean transmitted flux of the observations that we are comparing
to (the values of the mean transmitted flux are taken from \mmr).  
Renormalizing the optical
depth is equivalent to modifying the intensity of the ionizing
background, as long as the effect of collisional ionization and the
change in the gas temperature caused by the different
heating rate can be neglected
\citep[see][for a test that these effects are in fact
negligible]{tle98}.
The optical 
depth is then mapped to transmitted flux using $F=\exp(-\tau)$.

For each line of sight through the simulation (parallel to one of the three 
axes), we estimate the effects of 
continuum fitting by defining the maximum transmitted flux along the line of
sight to be
the continuum flux, $F_c$, and dividing the flux in all other pixels 
in the line by $F_c$.
We map the 288 cells along a line of 
sight onto smaller cells, with their size chosen to match the 
observations that we want to compare to.
Finally we convolve the spectra with the instrumental resolution of 
$6.6 \kms$, and
we add Gaussian noise to each cell with a flux dependent 
dispersion $n(F)$, which is
taken from \mmr.

\subsubsection{The $T-\Db$ Relation in the Simulation}

  Before we describe the method we shall use to measure the
temperature, it will be useful to examine the temperature-density
relation in the simulation.
In this paper we parameterize the mean temperature-density relation
(hereafter referred to as the \TDr) as
a power-law, $T=T_0 \Db^{\gmo}$, where the gas over-density is
$\Db\equiv \rho_g/\bar{\rho}_g$, for the purpose of measuring this
relation from the data. Although the \TDr
naturally approaches a power-law form with $\gmo \simeq 0.6$
when the thermal evolution
is determined by photoionization heating and adiabatic expansion alone
\citep{hg97}, in general it deviates significantly from a power-law.
\begin{figure}
\plotone{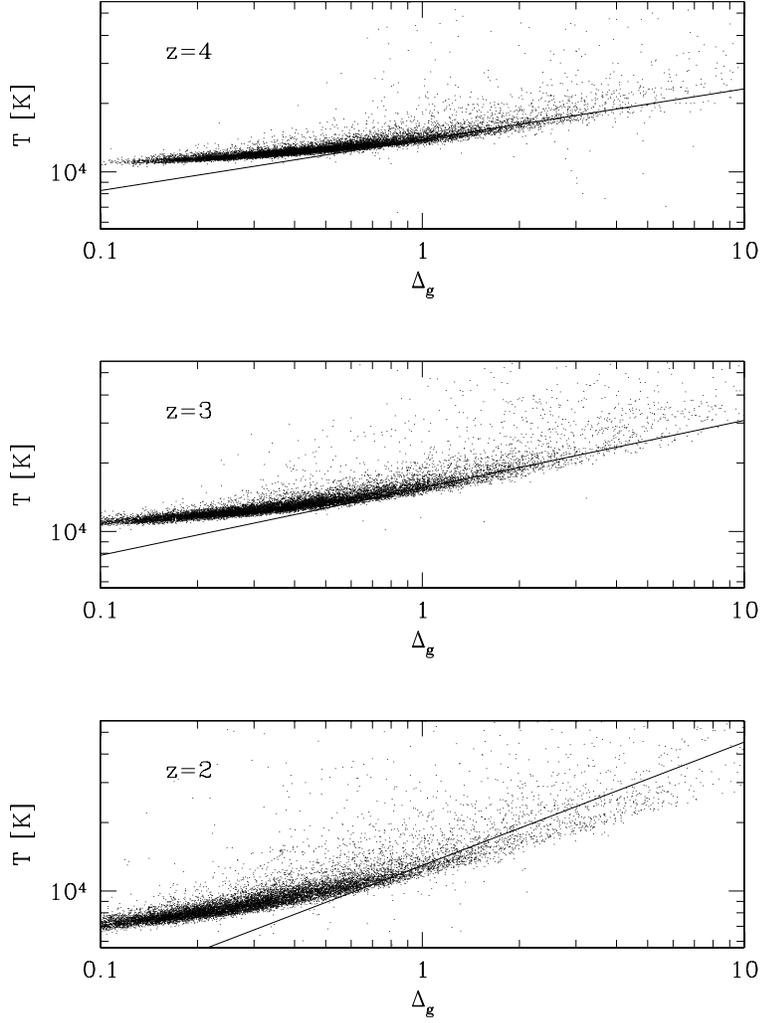}
\caption{Temperature vs. density for random points in real space
from the simulation. The lines are minimum absolute deviation
power-law fits to each set of points in the restricted range
$1<\Db<2$; the fitted parameters are given in Table \ref{basesimTdbtab}.
}
\label{TvsDbscatter}
\end{figure}
Figure \ref{TvsDbscatter} shows scatter plots of T versus
$\Db$ for the three redshift outputs of the simulation.  
The solid lines are power-law fits to the range 
$1<\Db<2$, to show that the
$T-\Db$ relation in the simulation is only roughly consistent with
a power law. There is a substantial dispersion of the temperature at
a given density, and the relation between the mean temperature and the
density deviates from a power-law. For example,
using the $z=3$ simulation output, the best power-law fits in the
restricted ranges of density $\Db=$(0-1, 1-2, 2-3) yield
$\gmo=$(0.15, 0.30, 0.39), with a mean fractional 
temperature deviation around the fits equal to (4\%, 11\%, 18\%).
The dispersion is due to
shock-heating and to the variable expansion or contraction histories of
the gas at a fixed density. The reionization of \heii occurs near
$z\sim 3$ in this simulation, heating the low-density gas to a
temperature that is nearly constant with density.
The parameters of the power-law fits to all three simulation outputs, in
the range $\Db=$(1-2), are given in Table \ref{basesimTdbtab}. 

  Because of this deviation from a power-law form of the \TDr,
a measurement of the temperature $T_0$ and power-law index $\gamma$
should be understood only as an approximation to the true mean
\TDr, near
the effective density at which the measurement is made, and this
effective density needs to specified. We use the following form to
present our results in \S 6:
\begin{equation}
T=T_\star (\Db/\Delta_\star)^{\gmo}~,
\end{equation}
where $\Delta_\star$ is chosen so that the error bars on 
$T_\star$ and $\gmo$ are uncorrelated.

\subsubsection{Definition of Temperature and Density at Points 
in Spectra}

  Each pixel in a spectrum receives optical depth contributions from an
extended stretch of real space along the line of sight, so no unique 
temperature or density can be associated with the pixel. However, we can
define the temperature and the gas density at pixels in spectra to be
the optical depth-weighted average over the temperature and the density
of all the gas that contributes to absorption in the pixel. This
definition will be used later to assign a gas temperature to any
identified absorption line, which we will define to be equal to the
temperature of the central pixel.
Table \ref{basesimTdbtab} shows fits to the \TDr
for pixels, restricted over the same density range $1 < \Db < 2$: we see
that the mean relation is almost the same as for random points in space.
This is true in spite of a larger difference in the density
distribution; for example, the median $\Db$ is $(0.47, 0.39, 0.31)$ 
for
random points at $z=(4, 3, 2)$, and is $(0.52, 0.47, 0.41)$ for spectral
pixels. This difference in the median density is caused by the thermal
broadening and velocity dispersion in absorbers, which spread high
density regions out into low density regions.

\section{FITTING METHOD}

In this section we describe the procedure that we use to identify and
fit absorption lines.
The temperature measurement is based on the identification of
absorption lines that can be adequately fit by a single Gaussian in
optical depth, over a certain interval around a point where
the optical depth is maximum; the narrowest widths among the lines
that can be fitted in this way will give us the gas temperature.
Absorption systems that cannot be fitted by a single Gaussian must be
broadened by a non-Gaussian distribution of the fluid velocity, and can
therefore be discarded for the purpose of measuring the gas
temperature. In contrast to the standard Voigt profile fitting approach,
we make no attempt to fit the entire spectrum by superposing many
absorption lines. Instead, we fit only small regions around minima of
the transmitted flux, each one with a single Gaussian absorber.
We therefore have a constant number of parameters to fit for each
absorption line, making the algorithm simple, unambiguous, and fast.

Before we explain the procedure for identifying and fitting lines, it is 
useful to understand qualitatively what the results will 
look like. 
\begin{figure}
\plotone{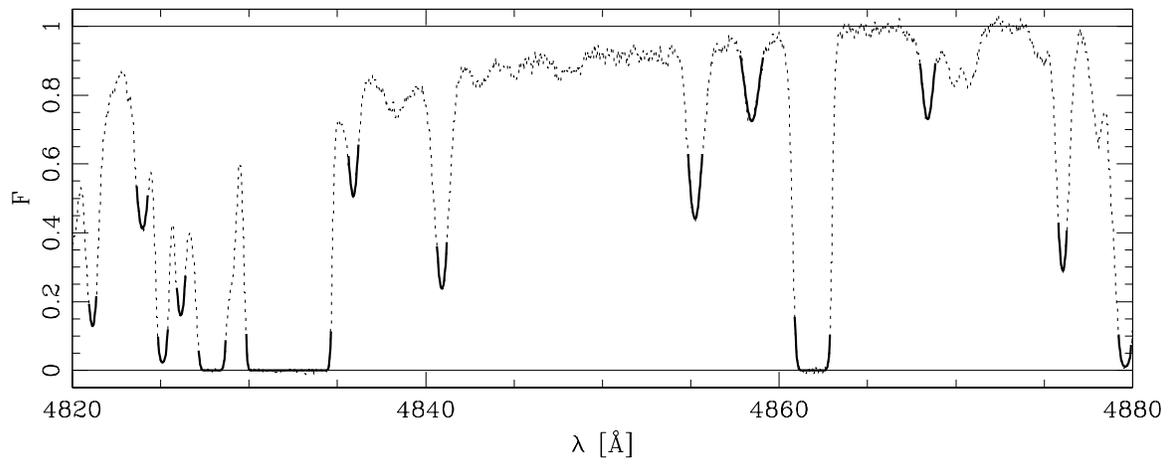}
\caption{The dotted line shows a piece of an observed spectrum.
The solid lines show the fitting results.
}
\label{fittedspec}
\end{figure}
Figure \ref{fittedspec} shows a section of the spectrum of Q1422, with 
the transmitted flux indicated by the dotted line, and the fitting solutions
indicated by the solid lines.  
Our method has selected all statistically significant maxima in optical depth
that are well fitted by a single Gaussian and identified them as absorption
lines and has discarded the rest.
The fits are done over the regions indicated by the solid lines.  
We will return to this figure once we
have described the process used to fit the lines.

  Our fitting method consists of taking each pixel in the spectrum as
a candidate for containing the center of an absorption line. After
requiring several conditions and eliminating most of the pixels as
candidates, a final list of absorption lines is obtained, each one having
a fitting window where a fit to the three parameters of the line (line
center, central optical depth, and line width) is performed. None of the
fitting windows from adjacent lines can overlap in the final list, as
seen in the example of Figure \ref{fittedspec}.

  The first operation is to determine an integer window width, $W$, which
sets the fitting region around each pixel $P$, going from $P-W$ to $P+W$.
The width $W$ is the smallest one for which the following condition
is obeyed, which ensures that there is a significant decline of
the flux from the average value at the edges of the fitting window to
the line center:
\begin{equation}
\frac{1}{2}\left[F\left(P+W\right)+F\left(P-W\right)\right]-F\left(P\right)>
E_d~\sigma\left(P,P\pm W\right)~,
\label{Edcondit}
\end{equation}
where 
\begin{equation}
\sigma\left(P,P\pm W\right) \equiv \left[\sigma^2\left(P\right)+
\frac{1}{4}\left(
\sigma^2\left(P+W\right)+
\sigma^2\left(P-W\right)\right)\right]^{1/2}~,
\end{equation} 
$\sigma(P)$ is the noise at pixel $P$, and
$E_d$ is the first parameter of the fitting algorithm.
$E_d$ is the number of ``$\sigma$'' significance required of the
flux decrease.
 
  There are of course some pixels where the condition
in equation (\ref{Edcondit}) is never obeyed for any
width. In practice, the width $W$ is increased only up to some maximum
value $W_{max}$ before the pixel is discarded as a candidate for a line
center. This maximum width is chosen to be large enough so that it does
not affect the results of the algorithm.
In addition, the minimum value of $W$ is set to 2 pixels (so that the fitting
region has at least 5 pixels). Some additional parameters are
used in the algorithm to expedite the elimination of pixels as candidates for
line fits, increasing
the speed of the code without affecting the final result; these details
are described in the Appendix.
 
  For each pixel $P$ where a fitting width $W$ has been determined in
this way, the line fit is performed by $\chi^2$ minimization. We fit the
following profile to the flux within the window:
\begin{equation}
F(v)=\exp\left[-\tau_c
\exp\left(-\frac{1}{2}\frac{\left(v-v_c\right)^2}{\sigma_b^2}\right)
\right]~.
\label{profeqn}
\end{equation}
Here, $v$ is the distance from the central pixel $P$. The three
parameters of the fit are $\tau_c$ (the optical depth at the line
center), $v_c$ (the location of the line center), and $\sigma_b$ (the
width of the line).
The parameter $v_c$ is constrained to lie within the pixel $P$, so that when
$P$ is not close to the center of the line, a good fit will not be obtained.
To account for the instrumental resolution, 
the function in equation (\ref{profeqn}) is convolved with a Gaussian filter 
of width matching the resolution of the data. 
After the fit is performed, we
impose a goodness-of-fit requirement: the probability of exceeding 
by chance the value of $\chi^2$ for the best fit should be larger than
a certain value $P_0$, which is a second parameter of our method.
If the requirement is not satisfied, the pixel $P$ is discarded as a
candidate for including the center of an absorption line.
The two parameters $E_d$ and $P_0$ can be adjusted to optimize the
temperature measurement, based on tests using numerical simulations
that we will present in \S 4.

  Because we require $v_c$ to be within pixel $P$, acceptable fits to
absorption lines are only found in pixels that are indeed close to a
minimum of the flux, in an absorption feature that can be adequately
fitted to a single Gaussian within a certain window.
Typically, the list of
pixels where acceptable fits are found will include groups of a few
adjacent pixels around such minima. The final step of our algorithm is
to select among any group of pixels with accepted line fits that are
within their own fitting windows the one that yielded the best fit.
This produces the final list of absorption lines.

We can now understand the fitting example shown in Figure \ref{fittedspec}.
The widths of the fitting windows, shown by the solid lines, are set
by the value of $E_d$ (here, $E_d=12$) and the noise level (which in this case 
varies from $\sim 0.004$
in saturated pixels to $\sim 0.01$ at the continuum). The apparent maxima of 
absorption that have no corresponding fitted line do not
increase in flux enough at their edges to satisfy the requirement in
equation (\ref{Edcondit}).
The cluster of fitted
lines near $\lambda=4830$\AA$~$ demonstrates that our procedure does not 
automatically eliminate lines in blends, as long as they are clearly
distinct maxima.  Apparently, the requirement that the optical depth be
consistent with a Gaussian curve is easily fulfilled by the peaks 
of all of the significant absorption lines.

  Throughout this paper, we shall be expressing all results concerning
the line widths in terms of the equivalent temperature, denoted as $B$,
when the line width is assumed to be due to thermal broadening:
$B \equiv 10000~(\sigma_b/9.09\kms)^2$ K. This allows for an easier
comparison of the results of numerical simulations and observations.
Note that line widths have usually been presented in the literature in
terms of the Doppler parameter, $b=2^{1/2} \sigma_b$.

  An important property of this algorithm is that the distribution of
line widths it measures should converge to a fixed answer as the
signal-to-noise ratio of the observations is increased. In the limit of
negligible noise and pixel size, and perfect resolution, every true
minimum of the transmitted flux in the
spectrum should be identified, and the fitted line width should
reflect the second derivative around the minimum, because the size of
the fitting window around each minimum should be very small. Moreover,
the second derivative around minima is a physically well motivated
quantity to obtain the gas temperature. In contrast, the Voigt profile
fitting method does not converge at high signal-to-noise ratio because
the number of blends assumed in an absorption system will change.
Whereas the Voigt profile method attempts to fit the entire spectrum
by superposing lines, our new method fits only small regions around
the minima of transmitted flux as arising from a single absorber.

\section{APPLICATION OF THE PROFILE FITTER TO THE NUMERICAL SIMULATION}

\subsection{Detailed Example of Fitting the Simulated Spectra}

We now apply our method to 1500 randomly selected lines of sight
through the simulation output at $z=3$,
with the mean flux decrement, noise level and pixel
size set to match the observations at $\bar{z}=3$
(see \mmr). We first set the two parameters of the
line-fitting algorithm to $E_d=12$ and $P_0=0.01$ (we shall analyze
the optimal values of these parameters in \S 4.2). A total of
6378 lines are identified and successfully fitted. For each
absorption line we obtain four quantities: the optical depth at the line
center, $\tau_c$; the Doppler parameter converted to temperature units,
$B$; the optical-depth-weighted temperature at the central pixel, $T$;
and the optical-depth-weighted gas density at the central pixel,
$\Db$.  

\begin{figure}
\plottwo{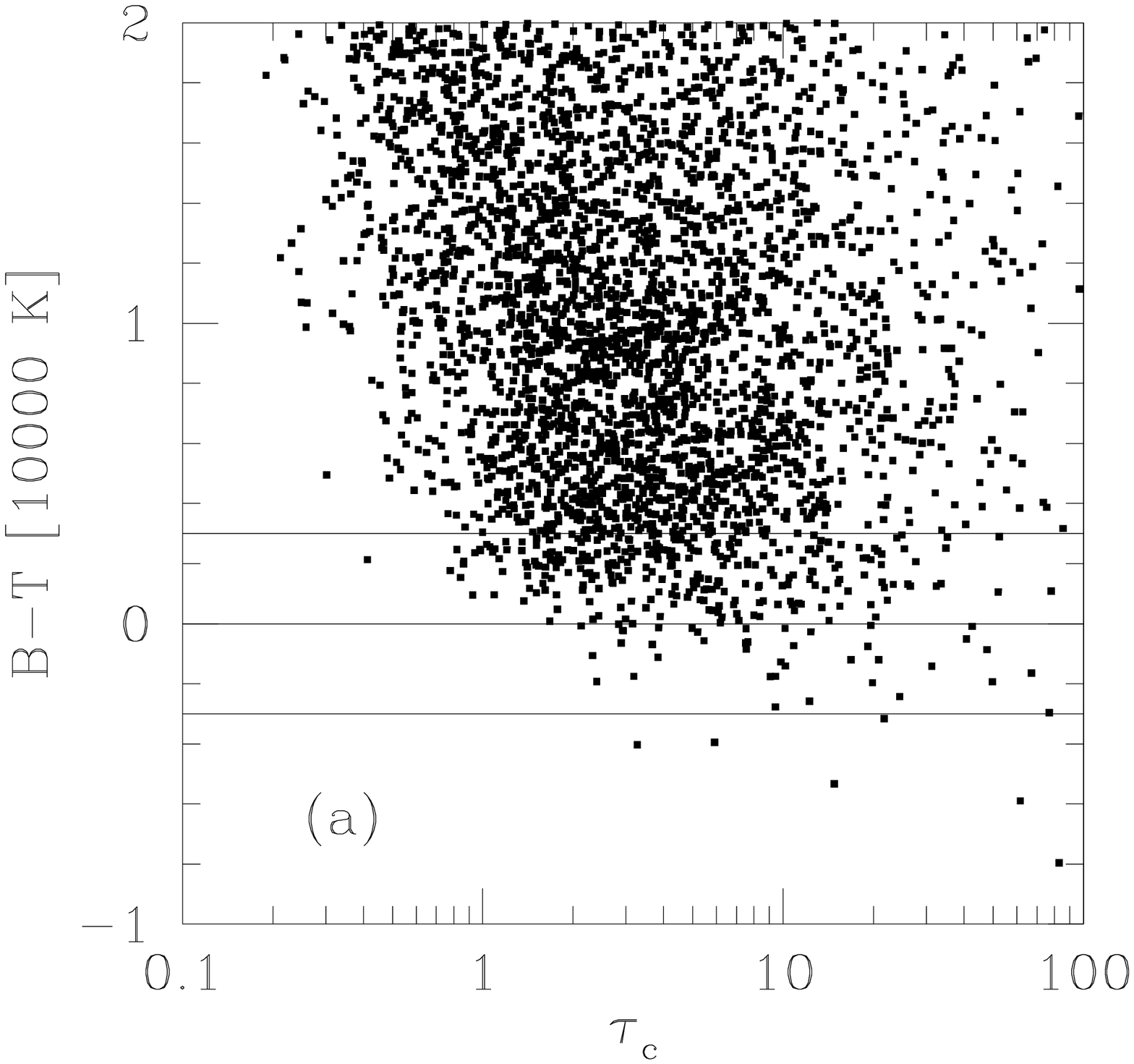}{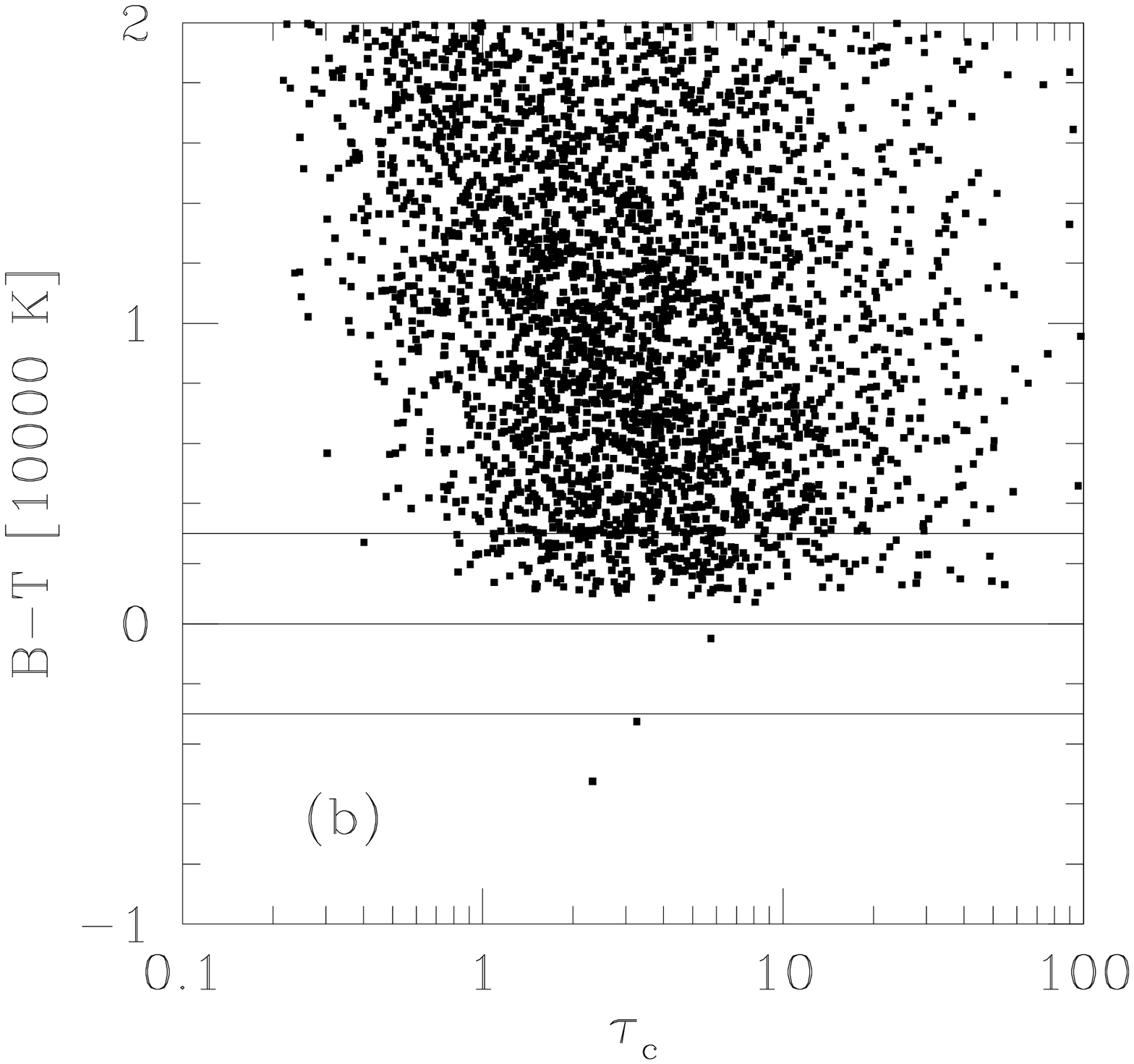}
\caption{The difference, $B-T$, between the temperature estimated
from the profile width and the 
true temperature at the profile center, plotted against the central
optical depth $\tau_c$, for absorption lines in the simulated spectra
at $z=3$. Noise at the same level as the observations at $\bar z = 3$
analyzed in \S 6 is included in (a), and not included in (b).
}
\label{DTscatter1}
\end{figure}
\clearpage
  Figure \ref{DTscatter1}(a) shows $B-T$ vs. $\tau_c$ for all 6378
fitted lines. The solid lines at $B-T=0$ K and $B-T=\pm 3000$ K are to
guide the eye in evaluating the contribution of non-thermal broadening
to $B$. The noise that has been added to the simulated spectra
(see \S 2.2.1) is responsible for most of the absorption lines with
$B<T$. In Figure \ref{DTscatter1}(b) we show the fitted lines from the
same set of spectra as Figure \ref{DTscatter1}(a), but without adding
noise (although the noise level that each pixel should have
was still used to weight
the $\chi^2$ fits); the small number of lines
that had $B<T$ in the presence of noise have been almost entirely eliminated 
(the few remaining lines with $B<T$ arise because the optical-depth-averaged
temperature at the central pixel of a line can be skewed by a contribution
from very hot gas).

Figure \ref{DTscatter1} demonstrates that the contribution to $B$ from
fluid motions is generally large, and therefore the Doppler parameter
of an individual line will usually overestimate the gas temperature.
For $\tau_c \gtrsim 1$, the distribution of $B-T$ extends to
values as low as $\sim 1000$ K, implying that the lower cutoff in the
$B$ distribution should provide a good measure of the median gas
temperature at a given $\tau_c$ [assuming that the lowest gas
temperatures do not extend very much below the median value, as is true
in the simulation (Fig. 1)]. However, for weak lines ($\tau_c < 1$),
thermal broadening is never clearly dominating, preventing a
model-independent estimate of the gas temperature at the correspondingly
low gas densities. The reason why the breadth of weak lines is always
dominated by motions is that all the low density gas has not turned
around from the Hubble expansion.  
\begin{figure}
\plotone{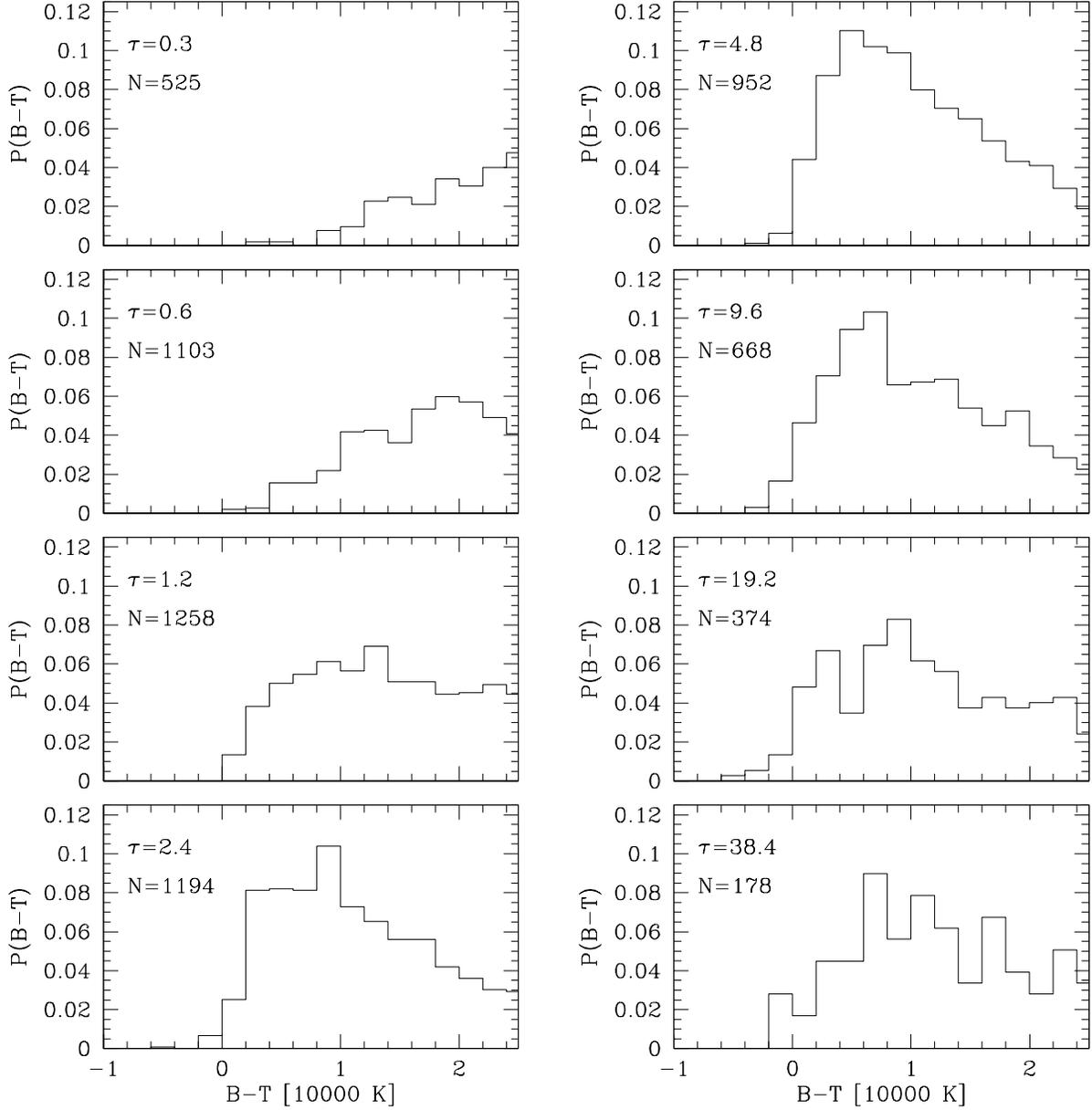}
\caption{Distribution of the non-thermal broadening contribution
to the line widths, $B-T$, at eight bins of the optical depth at the
line center, $\tau_c$. $N$ is the total number of lines in each 
$\tau_c$ bin.
}
\label{DThist}
\end{figure}

The eight histograms in Figure \ref{DThist}, constructed from the same
fitted lines as Figure \ref{DTscatter1}(a), show more quantitatively
the cutoff on the $B-T$ distribution near $B-T=0$, for different optical
depths.  In the 
$\tau \sim 0.3$ and $\tau \sim 0.6$ panels of the Figure, the problem
of estimating the temperature of the lower density gas is clearly seen.
Over the range $1\lesssim\tau\lesssim 20$, the distribution of $B-T$ is
the desirable one for our temperature measurement: there are many lines
near $B-T=0$, but very few with $B<T$.

The sharpness of the cutoff in Figure \ref{DThist} should of course be
degraded in the observable $B$ distribution, owing to the scatter in
the temperatures of the lines.
\begin{figure}
\plotone{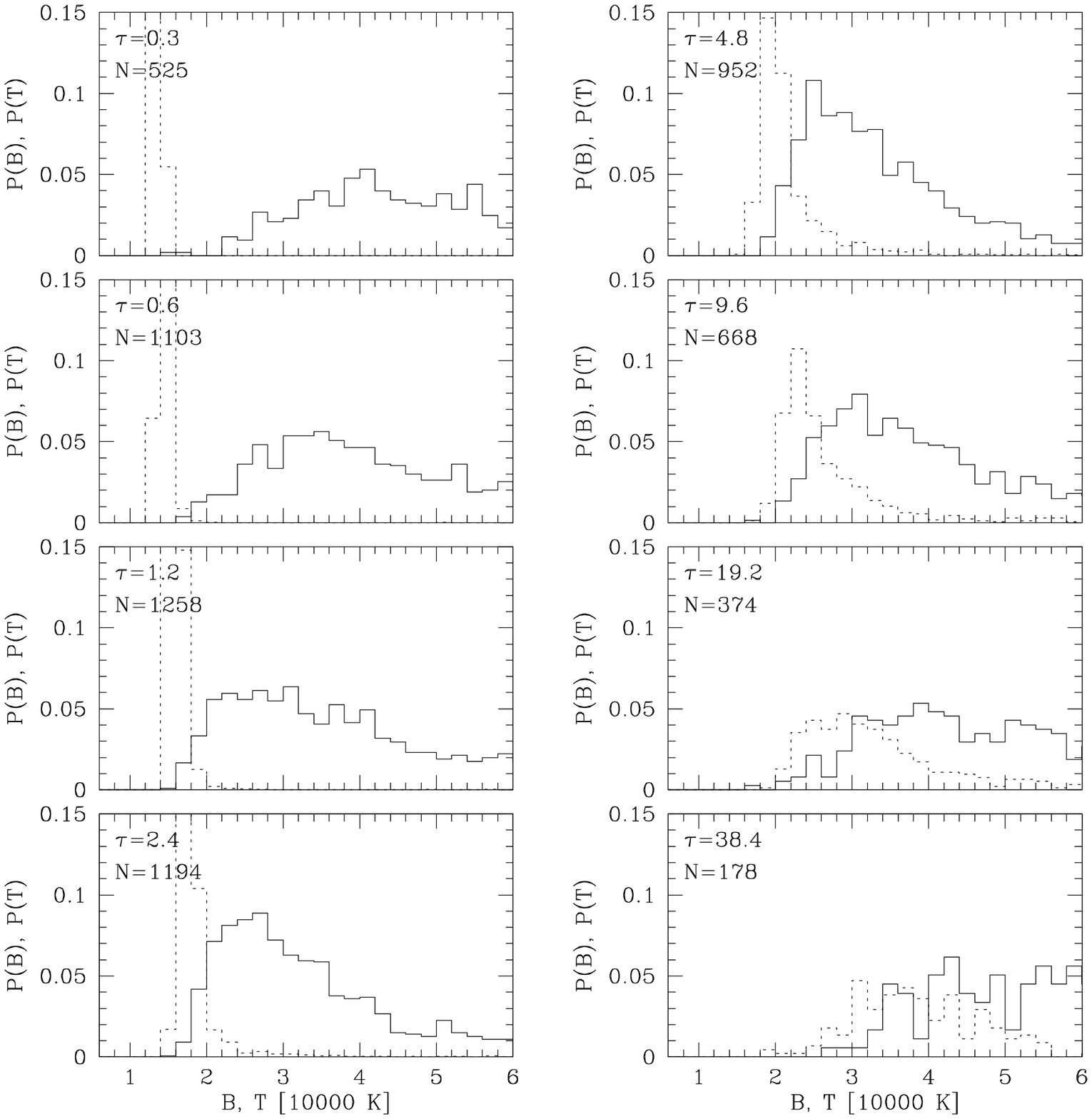}
\caption{The solid lines show the histograms of $B$ and the
dotted lines show the histograms of $T$.  
The fitted lines used here are the same ones shown in the
previous several Figures (described in \S 4.1).
The probabilities in the
$T$ histograms are all reduced by a factor of 2.5 to facilitate
visual comparisons.
}
\label{tThThists}
\end{figure}
Figure \ref{tThThists} shows
histograms of $B$ along with histograms of $T$.
By observing the histogram of $B$ we would like to determine the 
median of $T$.  For $\tau_c>1$, the cutoff on the distribution of 
$B$ appears to coincide well with the peak of the distribution of 
$T$.  As we saw clearly in Figure \ref{DTscatter1}, the number of
lines with $B \simeq T$ decreases quickly for $\tau_c<1$.

It is interesting to note in Figure \ref{tThThists} the change in
the distribution of the true temperature in the simulation with
$\tau_c$. At low optical depth, the $T$ distribution is very narrow as a
result of the simplicity of the evolution history of gas at low density,
which generally expands peacefully in the voids, heated by
photoionization and cooled adiabatically by its expansion.
As the density increases, the heating history of the gas becomes
more heterogeneous.  The gas at higher densities is shock-heated 
more frequently and to a greater degree, 
and the evolution of the density itself since reionization (when the
initial temperature is set) is more highly variable.  For 
$\tau \gtrsim 10$ the range of gas temperatures increases, making the
interpretation of the cutoff on the $B$ distribution more ambiguous.

\subsection{Optimizing the Fitting Control Parameters}

The two parameters of the line fitter, $E_d$ and $P_0$, should be
optimized to give the best statistical error bars on the
location of the lower cutoff of the distribution of $B$, which
we will use to estimate the gas temperature. In order to do this
optimization, we define a quality measure,
$Q=(N_g-N_b) (N_g+N_b)^{-1/2}$,
where $N_g$ and $N_b$ are the number of fitted lines with
$0~{\rm K} < B-T < 3000$ K, and with $-3000~{\rm K} < B-T < 0$ K,
respectively. Basically, $Q$ is a measure of
the statistical significance of the increase in the number of lines
as the $B=T$ cutoff is crossed, computed by comparing the number of
lines in bins of width 3000K on each side of the cutoff. The larger
the value of $Q$, the more accurately we should be able to determine
the temperature.
For example, the set of fitted lines shown in Figures 
\ref{DTscatter1}-\ref{tThThists} (fitted using $E_d=12$ and 
$P_0 = 0.01$)
gives $N_g=281$ and $N_b=40$, resulting in $Q=13.5$.
When we remove the noise in
Figure \ref{DTscatter1}(b) we find 
$N_g=236$ and $N_b=1$, yielding $Q=15.3$. We use a bin width of
$3000$ K because the errors in the measured temperature from our
data will be of about this magnitude (see \S 6).

Table \ref{Qchanges} shows the $Q$ values for a broad range of 
values of $E_d$, and $P_0= 0.01$. We also list the $Q$ value using two
different values of $P_0$, removing the noise from the spectra,
removing the continuum fitting approximation, and using different random
seeds for the added noise. We find that the changes in $Q$ are usually
smaller than the changes that can result from simply using a different
set of random numbers for the noise that is added to the spectra.
We conclude that the precise values taken by
the parameters are not actually very important to the results.
We use $E_d=12$ and $P_0=0.01$ when we analyze spectra matching
the $\bar{z}=3$ observational properties in the rest of this 
paper.  From the range of $E_d$ with $Q$ close to its maximum, we chose 
the
smallest value of $E_d$, $E_d=12$, because we expect that increasing the 
number of accepted lines will make the analysis procedure more robust,
particularly the computation of the error bars.  

  We use two other redshift bins for the observational data
(see \S 2.1), $\bar{z}=2.4$ and $\bar{z}=3.9$. Because the mean flux
decrement, pixel size, and noise level are different in each bin, we 
determine a best value of $E_d$ separately for each.
Tests similar to the one in Table \ref{Qchanges}, using the 
mean flux decrement, noise level, and pixel size matching the 
observations in the high and low $\bar{z}$ bins,
show that $E_d=9$ and $E_d=8$ are the best values to use when 
analyzing data in the 
$\bar{z}=2.4$ and $\bar{z}=3.9$ bins, respectively,
although the values of $Q$ obtained are only weakly sensitive to 
$E_d$ in each case.  
 We fix $P_0=0.01$ at all three redshifts,
because changing it does not significantly increase $Q$.

\section{TESTS OF THE DERIVATION OF THE IGM TEMPERATURE USING THE 
SIMULATION}

In this section we combine our line fitting method with elements of
the method of 
\citet{bm99} for constraining the temperature of the IGM
by measuring the lower cutoff on the distribution of $B$.  
First we define the cutoff and how we associate it with an estimated
temperature. Then we use the simulation to translate the observed 
temperature-optical depth relation into the desired 
\TDr. All the results in this Section are
obtained from the simulation. In \S 6 we present results for the
temperature measured from the observational data, using the method
described in this Section.

\subsection{Locating the Cutoff on the Distribution of $B$}

\citet{bm99} presented a useful technique for quantifying the
lower cutoff on a $B$ histogram like those in Figure \ref{tThThists}.
They smooth the histogram with a Gaussian filter and define the 
cutoff
to be the location of the maximum of the first derivative of the 
smoothed histogram.
This derivative is given by
\begin{equation}
\frac{dP}{dB}_i\propto \sum_{j=1}^N (B_j-B_i) 
\exp\left[-\frac{1}{2}\frac{\left(B_i-B_j\right)^2}
{\sigma_B^2}\right]~,
\label{difeq}
\end{equation}
where the $i$th bin has temperature $B_i$, 
$j$ is the label for $N$ individual fitted absorption lines with 
fitted temperature $B_j$, and $\sigma_B$ is the smoothing length.
\begin{figure}
\plotone{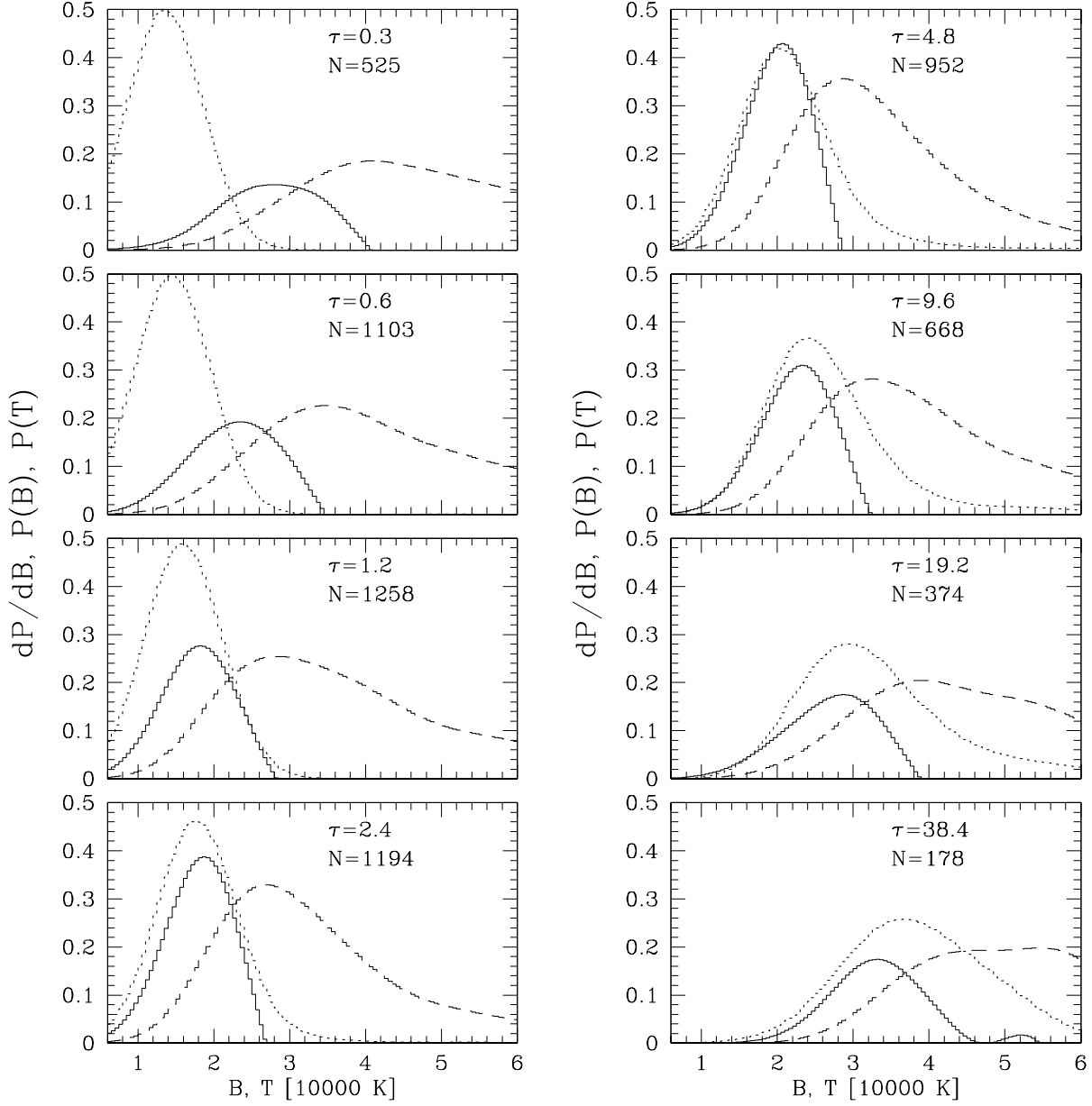}
\caption{The derivative of the smoothed histogram of $B$
({\it solid line}),
the smoothed histogram of $B$ ({\it dashed line}), and the
smoothed histogram of $T$ ({\it dotted line}), from the absorption
lines fitted in the $z=3$ simulation output with the
fitting control parameters $E_d=12$ and $P_0 = 0.01$.   
The smoothing length is $\sigma_B=5000$ K.  
The curves are arbitrarily normalized for clarity.
}
\label{smoothhist1}
\end{figure}
We plot in Figure \ref{smoothhist1}
the smooth histogram of $B$ and its derivative, as well as the
smoothed histogram of the gas temperature,
for $\sigma_B=5000$ K, using spectra from the $z=3$ simulation output
with the 
$\bar{z}=3$ observational noise, pixel size and mean flux decrement.

We define the Doppler parameter cutoff, $B_C(\tau_c)$, to be the value
of $B$ where $dP/dB$, given by equation (\ref{difeq}), is maximum,
for the optical depth bin labeled by $\tau_c$. Our estimate of the
gas temperature at optical depth $\tau_c$ is $B_C(\tau_c)$, after
applying a small correction that we describe in detail in the remainder
of this section.

\subsubsection{Error Bars on $B_C(\tau_c)$}

We compute error bars on the location
of the cutoff on the $B$ distribution ($B_C$) by bootstrap resampling
\citep{ptv92}.  We generate a bootstrap realization of the 
$B$ histogram, for an optical depth bin containing $N$ fitted 
lines, by
selecting $N$ lines at random from those in the bin (with 
replacement) and recomputing the histogram from the new set of 
lines.  The error on $B_C$ is given
by the dispersion in the $B_C$ values measured from many 
bootstrap realizations of the histogram. 

\subsection{Comparison Between $B_C(\tau_c)$ and the Temperatures in 
the Simulation}

  In this subsection we investigate the relationship between 
$B_C(\tau_c)$ and the physical temperature, $T$, of the absorbers with
central optical depth $\tau_c$.  We use spectra from the simulation,
where we know the optical-depth-weighted temperature and density, $T$
and $\Delta$, of the absorption lines. We match the simulated spectra to
the mean flux decrement, noise level and pixel size of the observations
at $\bar{z}=3$ (see \S 2.2.1). In \S 5.3 we show how the comparisons
change when the simulated spectra are matched to the properties of the
$\bar{z}=3.9$ and $\bar{z}=2.4$ observations.

\begin{figure}
\plotone{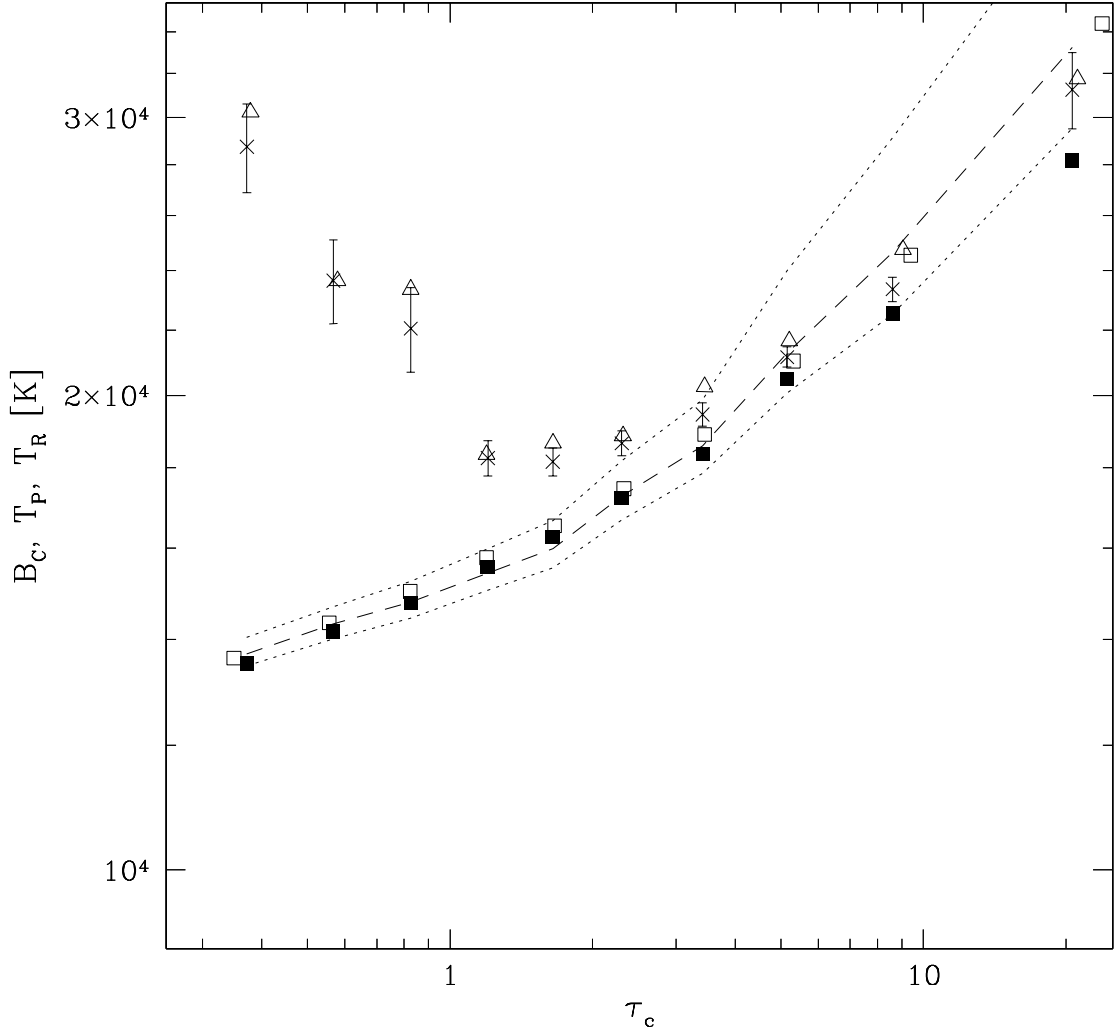}
\caption{
The crosses with error bars show $B_C(\tau_c)$, 
computed from the simulated spectra at $z=3$, and using
$\sigma_B=5000 \kms$. 
The open triangles show $B_C$ when the smoothing of the $B$ histogram is
reduced to $\sigma_B=3000$ K.  
The open squares are the median $T$ of all the absorption lines 
in each bin, while the filled squares are the median $T$ of lines
with $|B-B_C|<5000$K only.
The dashed line shows $T_R$,
the median value of $T$ for random cells in the simulation, 
and 
the dotted lines show the 25th and 75th percentiles.
}
\label{firstTest}
\end{figure}
Figure \ref{firstTest} explores the meaning of $B_C(\tau_c)$ in 
detail, using the $z=3$ simulation results.
We have separated all the fitted lines into 10 bins of the central
optical depth, $\tau_c$, choosing the bins so that each contains
an equal number of absorption lines. We smooth the $B$ histogram with
$\sigma_B= 5000 \kms$ to compute the cutoff temperature in every optical
depth bin. Then, we compute the median central optical depth and gas
temperature of the set of absorption lines that satisfy $|B-B_C|<5000$K,
in each optical depth bin. These sets of lines are the ones that actually
determine the location of the cutoff in the $B$ histogram, so we examine
first the relation between $B_C$ and their median temperature, which we
denote by $T_P$. The values of $B_C$ are shown as crosses with error bars.
\footnote{The error bars are from bootstrap realizations on 1500 lines
of sight. In reality, these error bars may be slightly underestimated
because the mean separation between 1500 lines of sight in the simulation
we use is only $\sim 60 \kms$, comparable to the flux correlation length
in the spectra (\mmr). Obtaining better statistics of the
theoretical prediction for $B_C$ would require a larger simulation.
However, for the analysis in the present paper the errors in the
determination of the temperature from the observations are much larger.}
The temperature $T_P$ is shown as the filled squares
(the errors on the temperature are much smaller than those on $B_C$).

The triangles in Figure \ref{firstTest} show the effect on the derived
cutoff, $B_C$, of reducing the histogram smoothing to $\sigma_B=3000$ K.
There is no important difference with the crosses (the error bars for
this smaller smoothing are similar to those on the crosses). By
experimenting with different values of $\sigma_B$, we have found that
$B_C$ is not affected if $\sigma_B$ is reduced, although the error bars
obtained are significantly larger when reducing it below
$\sigma_B=3000 \kms$. Increasing $\sigma_B$ beyond $5000 \kms$ leads to
an increase of $B_C$, because the smoothing is then larger than the
intrinsic sharpness of the cutoff in the $B$ distribution.
We therefore adopt $\sigma_B=5000$K from this point forward in the paper.  

The open squares show the effect of using {\it all} the absorption lines
in every optical depth bin to compute the median temperature and optical
depth, rather than using lines with $|B-B_C|<5000$K only. There is a
negligible difference in the derived temperature of the absorbers as a
function of $\tau_c$; the reason is that, as we have previously seen,
broader lines are by and large systems with higher fluid velocity
dispersions, but their gas temperatures are not significantly greater,
except at the highest optical depths where there is a slight difference
(the systematic shift to the right of the open squares relative to the
filled ones is due to a larger median optical depth of the broad lines
within each optical depth bin). In the rest of the paper, we always
compute the medians of any properties of the absorption lines
using only lines with $|B-B_C|<5000$K.

  So far, we have seen that the Doppler parameter cutoff $B_C$ provides
a good estimator for the gas temperature of absorption systems at a
given optical depth. Our ultimate goal, however, is to measure the
median gas temperature at a given gas density, for randomly selected
points in the IGM, which we shall henceforth refer to as $T_R$. The
peaks in absorption are at special locations, so their median
temperature, $T_P$, will generally not be exactly the same as $T_R$. 
To examine this question, we first compute the median density of the
absorption lines in each optical depth bin, and then we calculate the
median temperature of randomly selected points at this gas density.
The result is shown as the dashed line in Figure \ref{firstTest}; the
dotted lines give the 25 and 75 percentiles of the temperature
distribution at random points with the same gas density.
Comparing the dashed line to the filled squares, we see that the 
fitted absorption lines with relatively large optical depth typically have 
$T_P<T_R$, i.e., they are colder
than random fluid elements at the same density. We believe the reason
for this effect is the characteristic double-shock structure around the
absorbers \citep{cmo94}: the gas in the highest density tube along a
filament (or the highest density surface along a sheet) is located
between shocks, so it has been subject to less shock-heating than the
surrounding gas.

  Our method of analysis of the observational data in
\S 6 will automatically correct for this difference between the
temperature of the absorption lines and the temperature at random
points. This systematic difference may introduce a potential
uncertainty in the derivation of the gas temperature if it depends on
quantities like the resolution of the simulation, the cosmological
model that is assumed, or the heating at the reionization epoch.
However, the temperature difference between lines and random points
is negligible compared to the observational errors we will compute
for the temperature in \S 6, at least in the simulation we analyze here.

\subsubsection{Testing With Other $T-\Db$ Relations} 

We need to test the robustness of the finding that
$B_C(\tau_c)\simeq T_P(\tau_c)$ for $\tau_c\gtrsim 1$, which we have
established so far in tests on the $z=3$ simulation output.
We can change the \TDr that we are measuring
by simply using the $z=2$ or $z=4$ outputs from the simulation
(see Table \ref{basesimTdbtab}), still creating spectra with mean flux,
noise level, and pixel size 
matching the $\bar{z}=3$ observations, as described in \S 2.2.1.  In 
addition to the varying $T-\Db$ relations, 
these spectra have different amplitudes of fluctuations, and different
Hubble constants.

\begin{figure}
\plotone{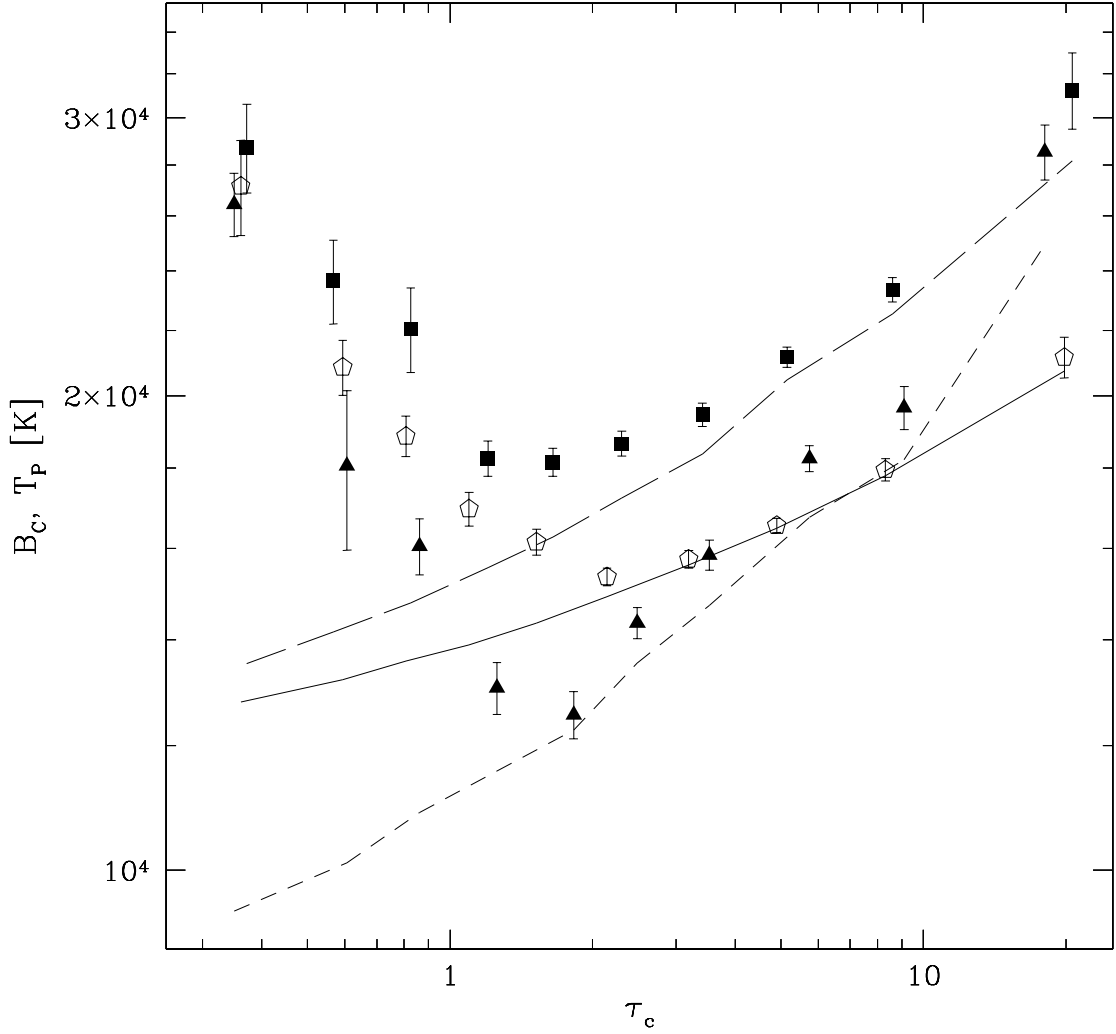}
\caption{({\it Pentagons, squares, triangles}):
$B_C$ for the $z=$(4, 3, 2) simulation output.
({\it Solid, long-dashed, short-dashed}) lines: $T_P$
(median temperature of fitted lines) for $z=$(4, 3, 2).
All simulated spectra have the mean flux decrement, pixel size and
noise level set to match the $\bar{z}=3.0$ observations.}
\label{Testdifredz3}
\end{figure}
Figure \ref{Testdifredz3} shows
the comparison between $B_C(\tau_c)$ and $T_P(\tau_c)$, using the 
$z=2$ and $z=4$ outputs of the simulation
in addition to $z=3$.  
The ({\it pentagons, squares, triangles}) 
show $B_C$ for the $z=$(4, 3, 2) simulation output,
and the ({\it solid, long-dashed,
short-dashed}) line show $T_P(\tau_c)$.
We see that $B_C$ traces the temperature changes
at the different redshifts extremely well, tracking the different
slopes at $z=2$ and $z=4$ perfectly, and matching the increased 
overall temperature at $z=3$.  All three redshift outputs show the
same strong increase in $B_C$ above the actual temperature for
optical depths below unity, corresponding to $\Db\sim 1$ in each 
case.  

\subsection{The $B_C - T$ Relation at Different Redshifts}

  We have seen that $B_C$ traces $T_P$ remarkably well for the mean
flux decrement and noise level of our observational data at $\bar z=3$.
We shall now verify that this is also true when the flux decrement and
noise is set to the values appropriate for the other two redshift bins
in which we separate the data, at $\bar z = 2.41$ and $\bar z = 3.89$.

  We first introduce a new type of figure that shows more clearly how
accurately $B_C(\tau_c)$ traces the gas temperature.  
\begin{figure}
\plotone{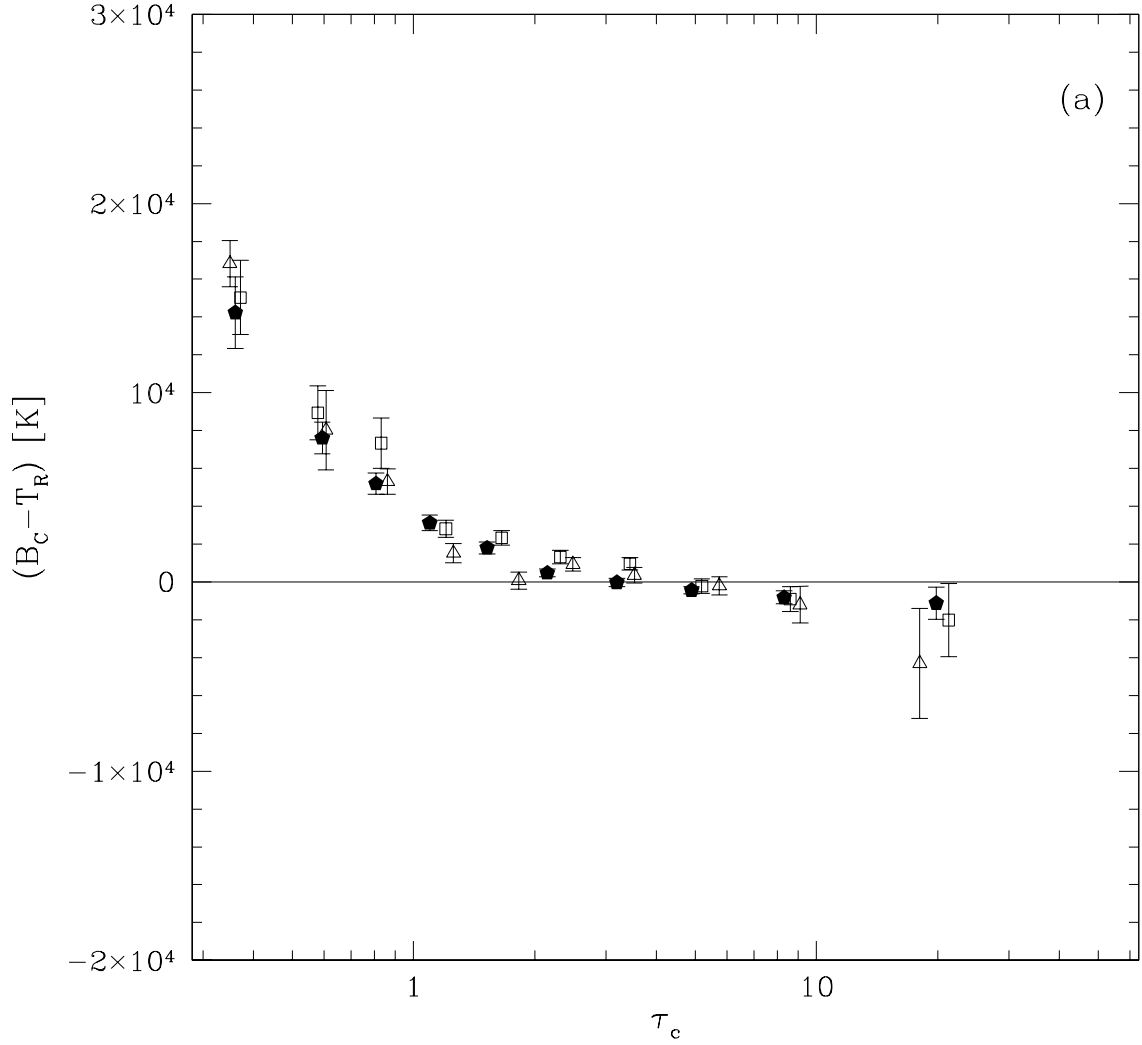}
\caption{({\it Pentagons, squares, triangles}):
$B_C-T_R$, at the $z=$(4, 3, 2) simulation output.
The mean flux decrement, pixel size, and noise level is set to
match the $\bar{z}=$(3.0, 2.4, 3.9) observations in panels (a, b, c).
}
\label{Testerrz3}
\end{figure}
\begin{figure}
\plotone{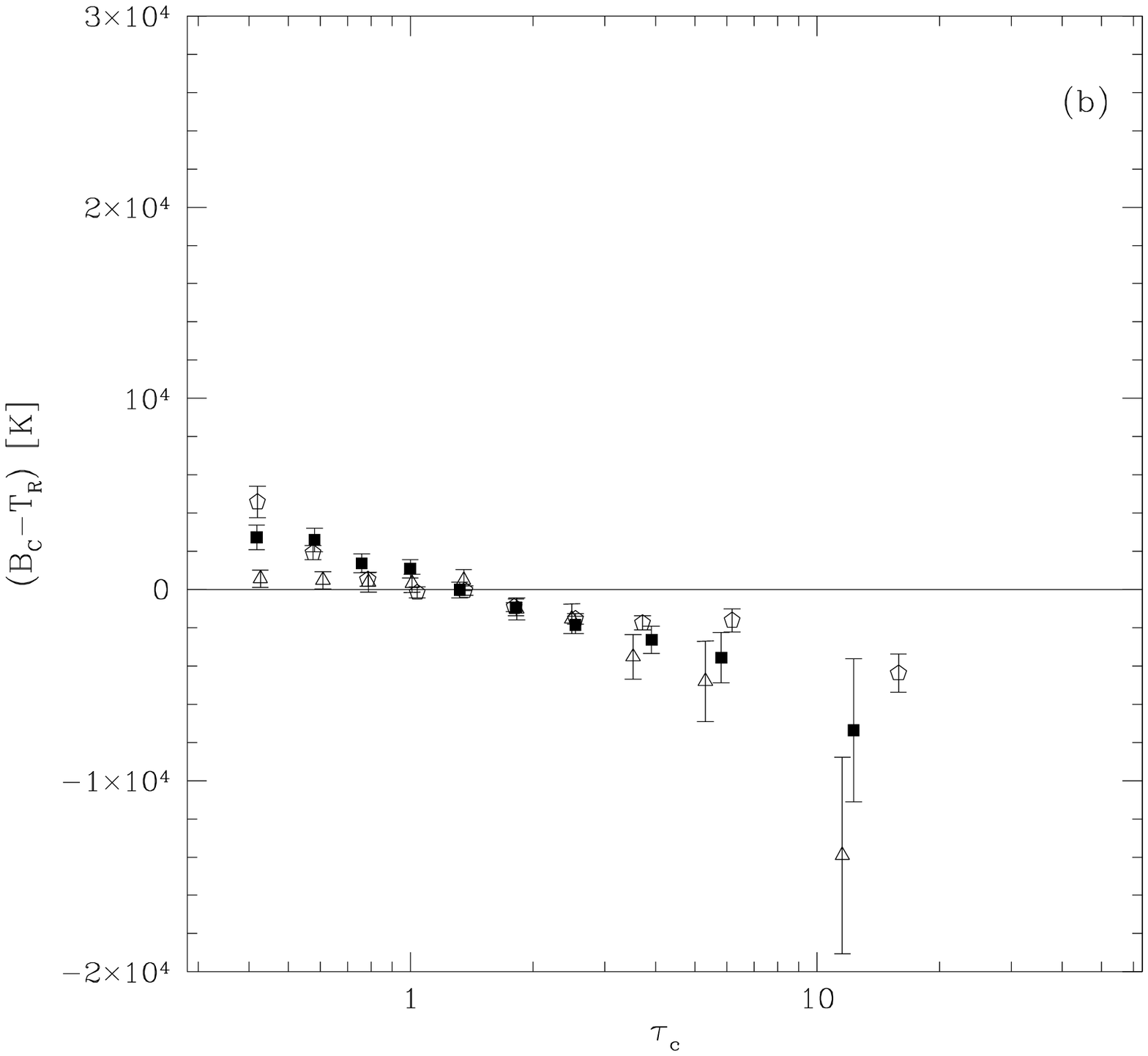}
\end{figure}
\begin{figure}
\plotone{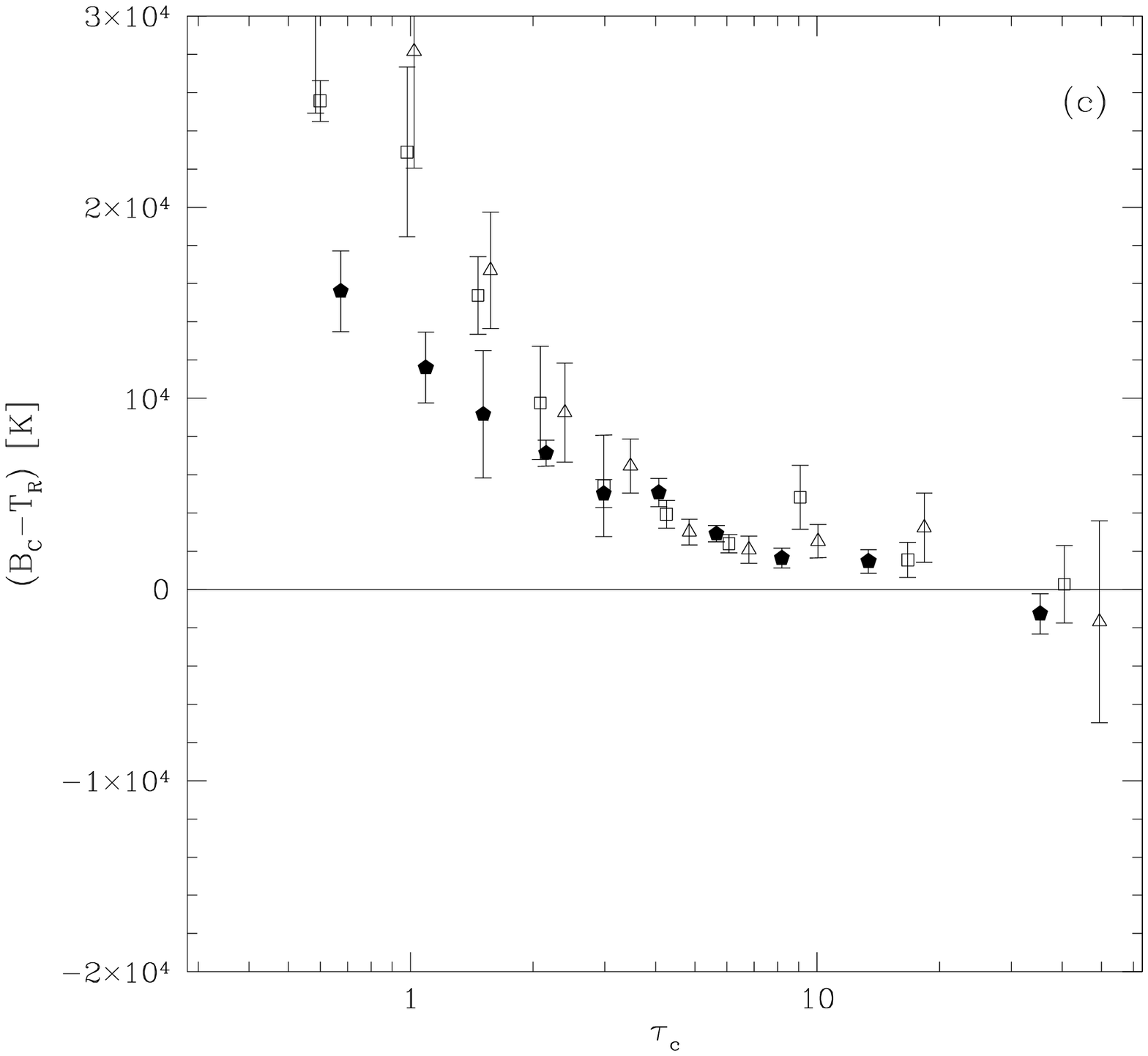}
\end{figure}
Figure \ref{Testerrz3}(a) shows $B_C - T_R$, where $T_R$ is the median
temperature of random points at the density of each optical depth bin
(recall that this density is defined as the median density of absorption
lines in each optical depth bin that have $|B-B_C|<5000$K). The
({\it pentagons, squares, triangles}) are obtained from the simulation
outputs at $z=$(4, 3, 2), and are the same points as in Figure
\ref{Testdifredz3} except that we have subtracted $T_R$. The temperature
is correctly traced by $B_C$ at $\tau_c \gtrsim 1$, and drops below
$B_C$ at lower optical depths in the same way, independently of which
simulation output we use. At $\tau_c > 10$, $B_C$ falls significantly
below the temperature in the $z=2$ output (this is seen clearly using more
optical depth bins). The reason is that the
temperature dispersion is higher at $z=2$ in the simulation, causing
$B_C$ to reflect the lower cutoff of the true temperature distribution.

  The results when we fix the mean flux decrement and noise to the
$\bar{z}=2.41$ redshift bin of the observational data 
(taken from \mmr) are shown in Figure \ref{Testerrz3}(b). Here,
we use
the parameters $E_d=9$ and $P_0=0.01$ for the line fitting algorithm
(see \S 3). The results are shown also using all three simulation
outputs. Again, we find the temperature is well traced by $B_C$, but
over a range of $\tau_c$ that has shifted to lower values.
This is a result of the decreased optical depth at fixed gas density.
Absorption systems that are primarily thermally broadened exist above
a gas density that is approximately constant at each redshift, but
the corresponding optical depth varies rapidly with redshift. At
high optical depths, an additional effect is important in changing the
degree to which $B_C$ traces the gas temperature: the increased
shock-heating at low redshifts (with increasing velocities of collapse)
implies a higher temperature dispersion, even at a fixed gas density.
Therefore, $B_C$ drops further below the median gas temperature as the
redshift decreases.

The result for $\bar z=3.9$ is shown in Figure \ref{Testerrz3}(c) (we
use $E_d=8$ at this redshift).
The Doppler parameter cutoff ($B_C$) now traces the temperature only at
high optical depth, $\tau_c \gtrsim 5$, for exactly the same reason:
a fixed gas density has shifted to a significantly higher optical
depth due to the increase in the mean flux decrement.

\subsection{Determining $T_R(\Db)$ Using $B_C(\tau_c)$}

  So far, we have seen that the Doppler parameter cutoff traces the gas
temperature over a reasonable range of optical depth. We have shown
this to be a consequence of the presence of some absorption lines that
are primarily thermally broadened, and of the small dispersion of the
gas temperature. We therefore expect that this relation between $B_C$
and $T_R$ will not be significantly changed depending on the model
adopted in the simulation, or when the numerical resolution is
increased.

  However, even over a restricted range of optical depth where $B_C$ and
$T_R$ are best matched, the simulation predicts a difference between
them, which we want to correct for when analyzing the observational data
(although this correction could be model dependent, and will need to be
compared with other simulations in future work). In addition,
we need to relate the central optical depth $\tau_c$ to the
optical-depth-weighted gas density of the absorber, $\Delta_b$, in order
to derive the \TDr of the gas from the observed
$B_C$ as a function of $\tau_c$. Given the limited amount of data that
we will analyze in this paper, it will be sufficient to
parameterize the \TDr by a power-law,
\begin{equation}
T=T_\star (\Db/\Delta_\star)^{\gmo}~.
\label{Tplequation}
\end{equation}
where $\Delta_\star$ is chosen to make the error on $T_\star$ and $\gmo$
uncorrelated. As discussed in \S 2.2.2, this power-law should be
understood only as a fit to the true relation, which should be more
complex. The power-law form will be adequate here given our error bars,
but larger sets of data might be used to detect deviations from a
power-law.

  In this subsection we develop the method to derive the parameters
$T_{\star}$ and $\gamma$, given the determination of $B_C$ at
different optical depths.

\subsubsection{Accounting for the Systematic Offsets $B_C - T_R$}

  To obtain an accurate estimate of the gas temperature from
observations, the systematic differences between $B_C$ and $T_R$ shown
in Figure \ref{Testerrz3} can be used to correct the observed $B_C$.
However, this correction may be dependent on the model, and this
dependence will not be known until a wide variety of additional
simulations are analyzed. We therefore use only absorption lines
over the range of $\tau_c$ where the offset between $B_C$ and $T_R$ is
small ($|B_C-T_R|<5000$K for all three simulation outputs). The following 
optical depth ranges will be used:
$0.41<\tau_c<5.4$ for $\bar{z}=2.4$, $1.0<\tau_c<19$ for $\bar{z}=3.0$,
and $3.8<\tau_c<47$ for $\bar{z}=3.9$.

  For each fitted line obtained from the observations (within the
accepted optical depth range), we determine a temperature correction at
the optical depth of the line by linearly interpolating from the two
adjacent points in Figure \ref{Testerrz3}. The set of points used
depends on the redshift bin of the observations and the simulation
output we choose for the analysis. The corrected line width is
$B'=B- \Delta T(\tau_c)$, where $B$ is the observed line width,
$\Delta T(\tau_c)=B_{C,S}(\tau_c)-T_{R,S}(\tau_c)$, $B_{C,S}(\tau_c)$ is
the cutoff of the $B$ distribution in the simulation, and
$T_{R,S}(\tau_c)$ is the median temperature in the simulation at random
points with gas density equal to the median density of the absorption
lines satisfying $|B-B_C(\tau_c)|<5000$K.  

\subsubsection{Translating $\tau_c$ into $\Db$}

  After the corrections just described, we have an estimate of the 
temperature as a function of $\tau_c$. What we need, however, is the
temperature as a function of $\Db$. This will be derived by using a
transformation from $\tau_c$ to $\Db$ that we obtain from the
simulation. This introduces an inevitable model dependence in our
measurement: the relation between $\tau_c$ and $\Db$, given a fixed flux
decrement, should essentially be subject to the same uncertainties that
appear in deriving the parameter $\mu \sim (\Omega_b h^2)^2/\Gamma/H(z)$
(where $\Gamma$ is the photoionization rate due to the cosmic
background) from the observed mean flux decrement 
\citep[see][, \mmr]{rms97}.

\begin{figure}
\plotone{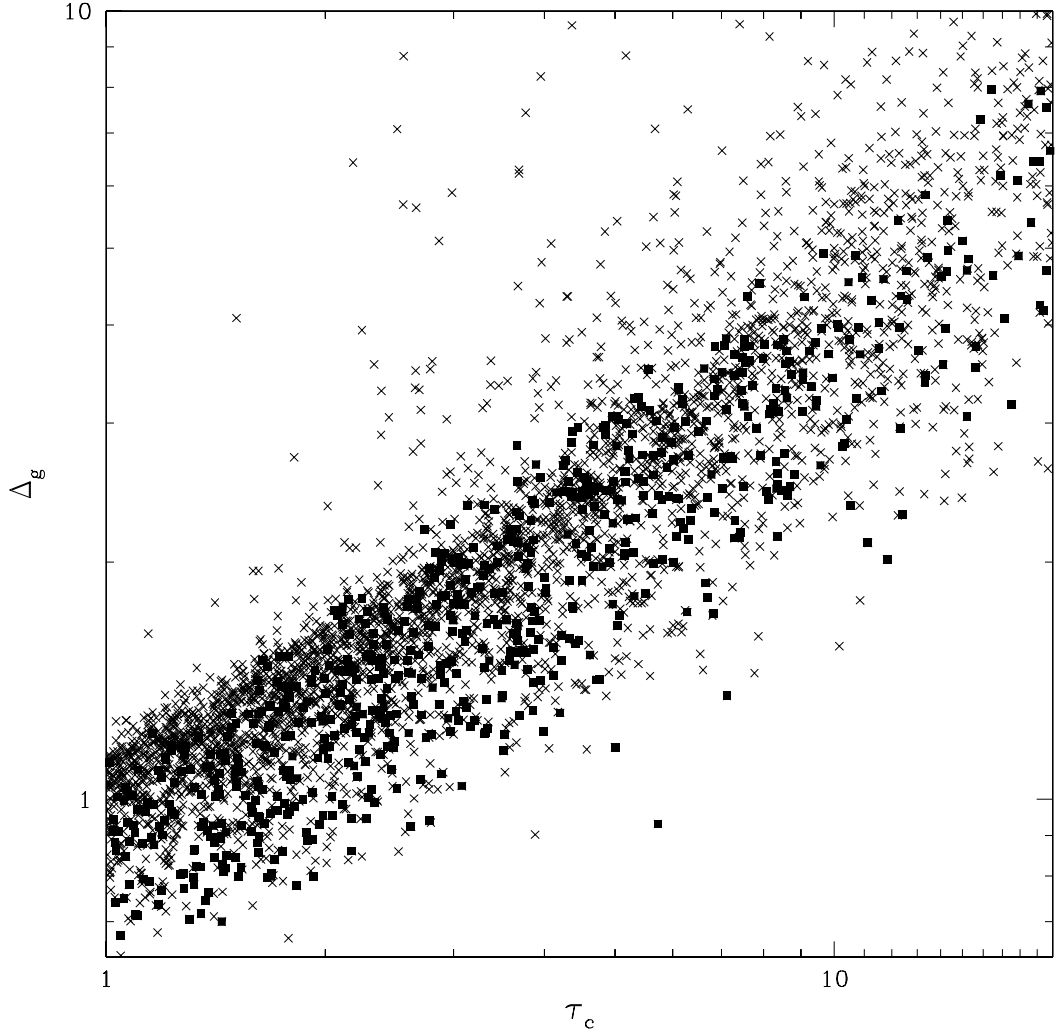}
\caption{$\Db$ vs. $\tau_c$ for a set of fitted absorption lines from the
$z=3$ simulation output.  Crosses mark all the lines, squares indicate the
lines with $|B-B_C(\tau_c)|<5000$K.
}
\label{Dbtaucor}
\end{figure}
Figure \ref{Dbtaucor} is a scatter plot of the density and optical
depth of the lines fitted from the $z=3$ simulation output (with
mean flux decrement matching the $z=3$ observations). The crosses show
$\Db$ vs. $\tau_c$ for all of the lines, while the filled squares show only
lines satisfying $|B-B_C(\tau_c)|<5000$K, where $B_C(\tau_c)$ was
determined for 10 optical depth bins as described earlier. The lines that
determine the $B$ cutoff (shown as squares) tend to have a higher
optical depth than other lines at the same density, because of their
lower velocity dispersion.

  To obtain the $\Db-\tau_c$ relation, we assign the median density
of the lines in each optical depth bin that we use to measure the
temperature (i.e., those with $|B-B_C(\tau_c)|<5000$K)
with the median $\tau_c$ of the same lines.  We use interpolation to
calculate the density corresponding to any value of $\tau_c$ for the
fitted lines in the observed spectra.

  One of the quantities affecting the $\tau_c - \Delta_b$ relationship
is the density-temperature relation itself, essentially because the
temperature affects the recombination coefficient, which then changes
the neutral fraction at a given density. In order for our determination
of the \TDr to be self-consistent, we need to change the
temperatures in the simulation so that they agree with the same
\TDr. We do this in the following way: after determining
a preliminary \TDr using the simulation with the true
temperatures, we modify the temperature in every cell of the simulation
using the formula
\begin{equation}
T_i'=T_i-T_\star (\Delta_i/\Delta_\star)^{\gmo}+
T_\star' (\Delta_i/\Delta_\star)^{(\gmo)'}~,
\end{equation}
where $T_i$ is the original temperature at cell $i$, $T_i'$ is the
modified temperature, $T_\star$ and $\gmo$ are the parameters of the
original \TDr of the simulation, and $T_\star'$ and
$(\gmo)'$ are the parameters of the new \TDr that we
wish to impose. We then iterate the application of this formula until
the modified \TDr of the simulation matches the one from the
observations. This modification of the temperatures in the simulation
causes only a small change in the derived \TDr (the
value of $T_0$ is modified by only $\sim 5$\%).

\subsubsection{Fitting for $\gmo$ Without Binning}

  We now describe the method we use to fit the parameters $T_\star$
and $\gmo$ to the values of $\Db$ and $B'$ of a set of fitted lines.
The simplest method would be to separate the lines in density bins,
measure the cutoff $B_C$ in every bin, and then fit the power-law
relation to the values of $B_C$ obtained at every bin. However, the
binning could result in a degradation of the measurement errors: at
least 50 lines are needed to obtain a reasonable estimate of the
cutoff $B_C$, and since our data yield only a few hundred lines for
each redshift bin, the binning in line density would need to be very
coarse.

  There is a simple solution to this binning problem if the cutoff
on the distribution of fitted lines, in the $B'-\Db$ plane, is
described by a power-law (recall that $B'$ is the width of each fitted
profile minus the expected non-thermal broadening at its optical depth,
as described in \S 5.4.1). After we have associated a gas density
$\Db$ with each fitted line and corrected their temperatures using the
simulation predictions, we rotate the absorption lines in the $B'-\Db$
plane for many different assumed values of $\gmo$, using the formula
\begin{equation}
B''=B'-T_\star (\Db/\Delta_\star)^{\gmo}~.
\label{rotateT}
\end{equation}
For each assumed $\gmo$, we apply the cutoff determination technique to
the $B''$ distribution (without density binning) to find a value for the 
temperature and a maximum value for $dP/dB$ [see eq.\ (\ref{difeq})]. As
$\gmo$ is varied, the best fit value of $\gmo$ is the one that results
in the maximum value of $dP/dB$, i.e., the sharpest cutoff on the $B''$
distribution.

  Note that the value of $T_\star$ used in equation (\ref{rotateT})
affects the size of the temperature changes in the rotation. We
therefore also iterate in the determination of $T_\star$ and $\gamma$ by
this procedure. In practice, the measurement of $T_\star$ is barely
affected by the rotation in equation (\ref{rotateT}), so a single
iteration is sufficient.

  Before presenting the results of applying our method to the
observations, it will be useful at this point to summarize the full
procedure we have described for measuring $T_\star$ and $\gmo$ from the
$B$ distribution of the fitted absorption lines at every redshift bin.
This consists of the following steps:\\
1.  Eliminate the absorption lines with $\tau_c$ outside the range
where the temperature measurement is expected to be effective.\\
2.  Correct the values of $B$ for all of the remaining lines
using the systematic offset predicted by the simulation 
(Figure \ref{Testerrz3}).\\
3.  Use the $\tau_c-\Db$ relation in the simulation to associate
a value of $\Db$ with each fitted line.\\
4.  Determine an initial estimate for $T_\star$ and $\gmo$ with the
method of fitting the power-law cutoff in the $B$ distribution as a
function of $\tau_c$ that was just described.\\
5.  Modify the \TDr in the simulation to
more closely approximate the relation measured from the
observations.\\
6.  Repeat steps 2-5 until the \TDr measured
from the observations matches the relation in the simulation.

\section{ANALYSIS OF THE OBSERVED SPECTRA}

  We now apply our line-fitting method to the observed spectra.
Table \ref{basicobsstats} lists, for each redshift bin, the redshift
range ($z_{min}$ to $z_{max}$) and mean redshift $\bar z$, the mean flux
decrement $\bar F$, the mean noise, the total number of pixels, the
total path length, the value of $E_d$ we use to fit the lines, the total
number of lines fitted, and the number of lines that we actually use to
measure the gas temperature in the range of central optical depth from
$\tau_{min}$ to $\tau_{max}$.
 
\begin{figure}
\plotone{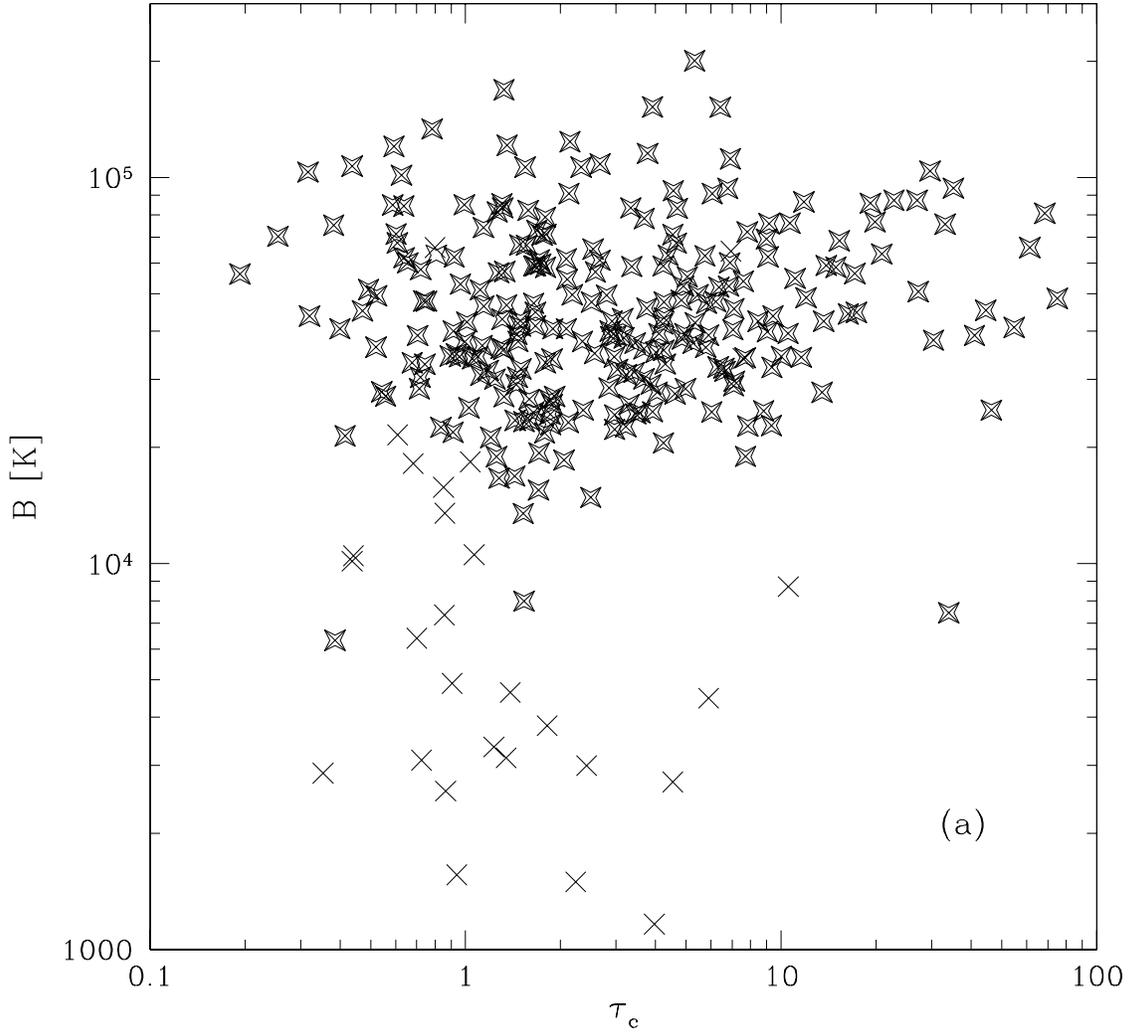}
\caption{(a, b, c) Fitted lines in the observed spectra
at $\bar{z}=$(3.9, 3.0, 2.4). Outlined crosses
are lines that are not in regions containing potential metal lines.
}
\label{z3obsscatter}
\end{figure}
\begin{figure}
\plotone{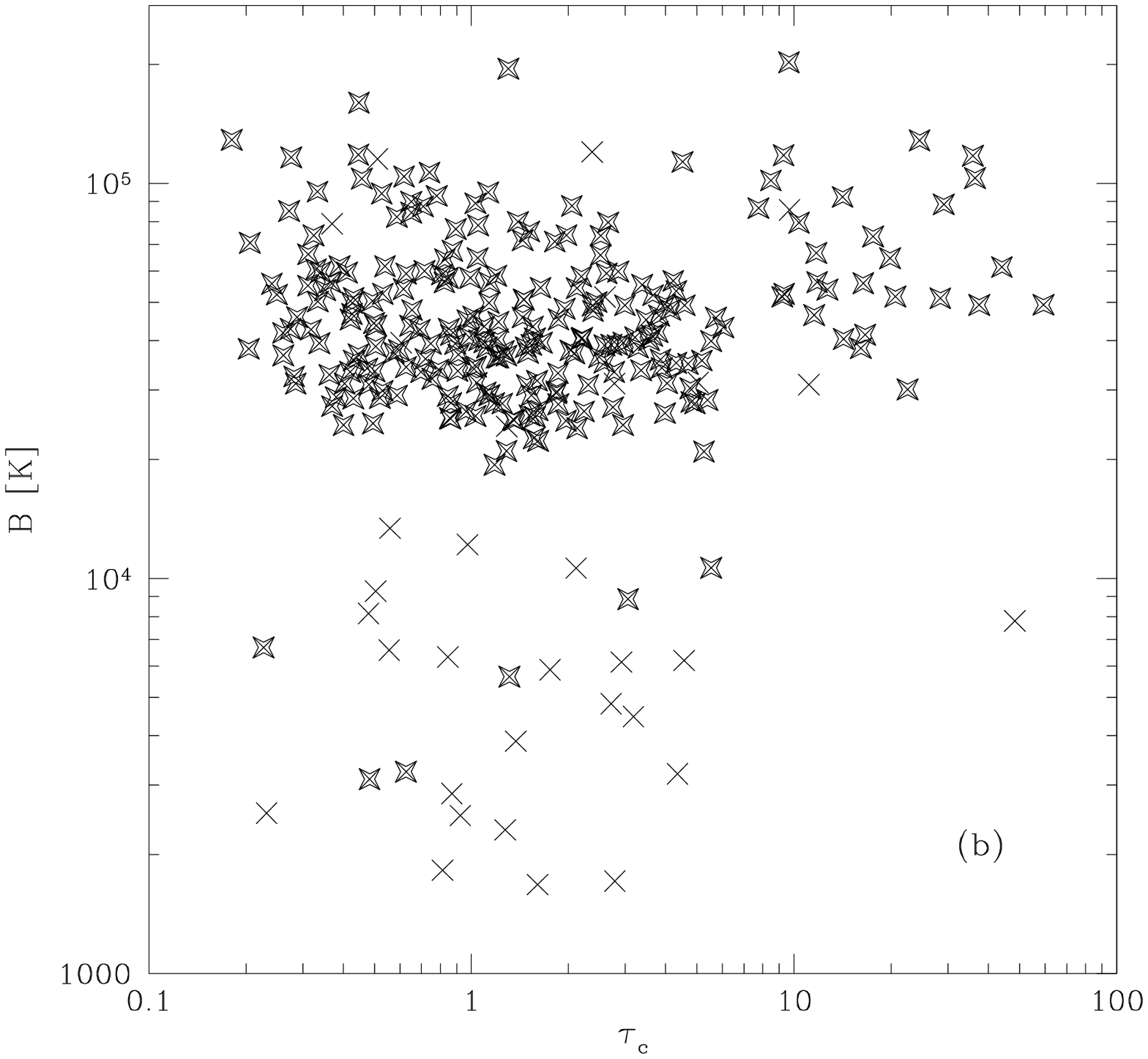}
\end{figure} 
\begin{figure}
\plotone{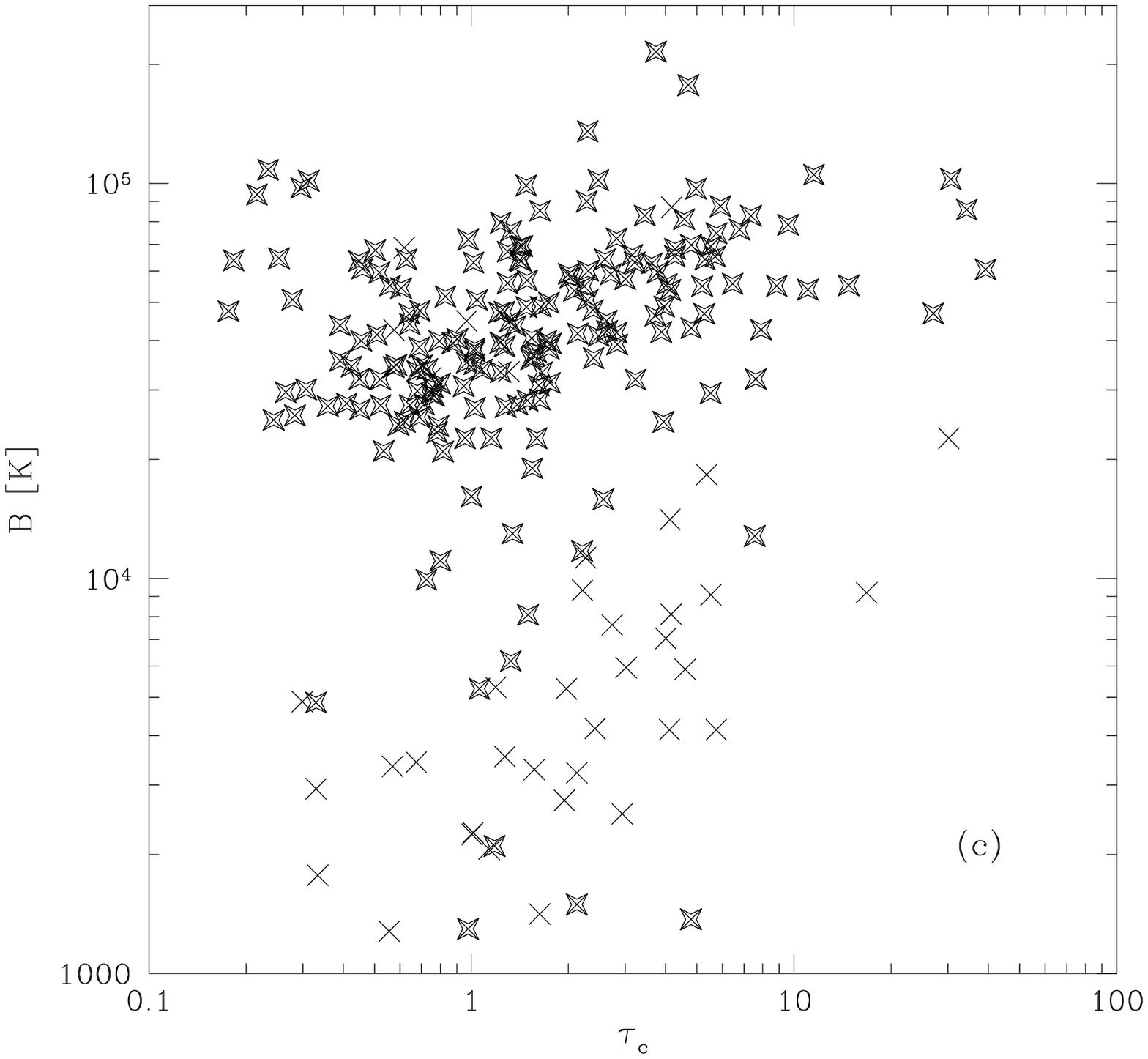}
\end{figure}
The parameters $(B,\tau_c)$ of all the fitted lines are shown in Figure
\ref{z3obsscatter}(a,b,c) for the redshift bins
$\bar{z}=$(3.9, 3.0, 2.4), as outlined crosses when the lines are not
in any of the regions suspected to include metal lines, and as simple
crosses when they are. For the quasar KP 77 (included in the redshift
bin $\bar z=2.4$), the analysis to identify potential metal lines was
not done, so all lines from this quasar are shown as outlined crosses.
The lower cutoff on the $B$ distribution is clearly visible to the eye
at all three redshifts, especially when the metal lines are ignored.  

\begin{figure}
\plotone{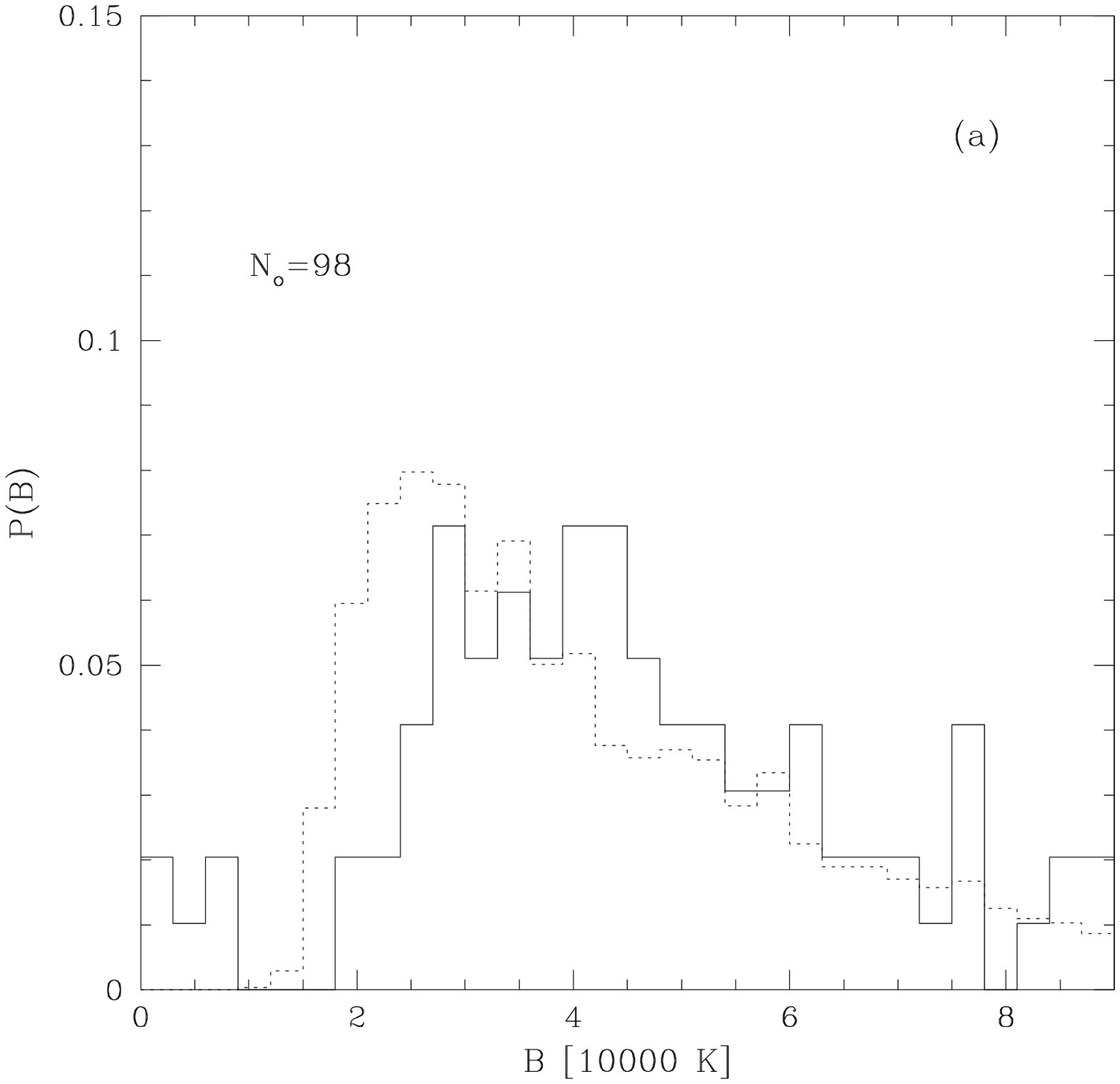}
\caption{Distribution of $B$ of fitted absorption lines.
(a) {\it Solid line:} observed lines at $\bar{z}=3.9$,
with $3.8<\tau_c<47$.
  {\it Dotted line:} Lines from simulation output at $z=4$,
with $3.8<\tau_c<47$.
(b) Same as (a) for the $\bar z = 3$ redshift bin and $z=3$ simulation
output, with $1.0<\tau_c<19$.
(c) Same as (a) for the $\bar z = 2.4$ redshift bin and $z=2$ simulation
output, with $0.41< \tau_c<5.4$.
$N_o$ is the total number of observed lines in each histogram.
}
\label{z3compsimobshist}
\end{figure}
\begin{figure}
\plotone{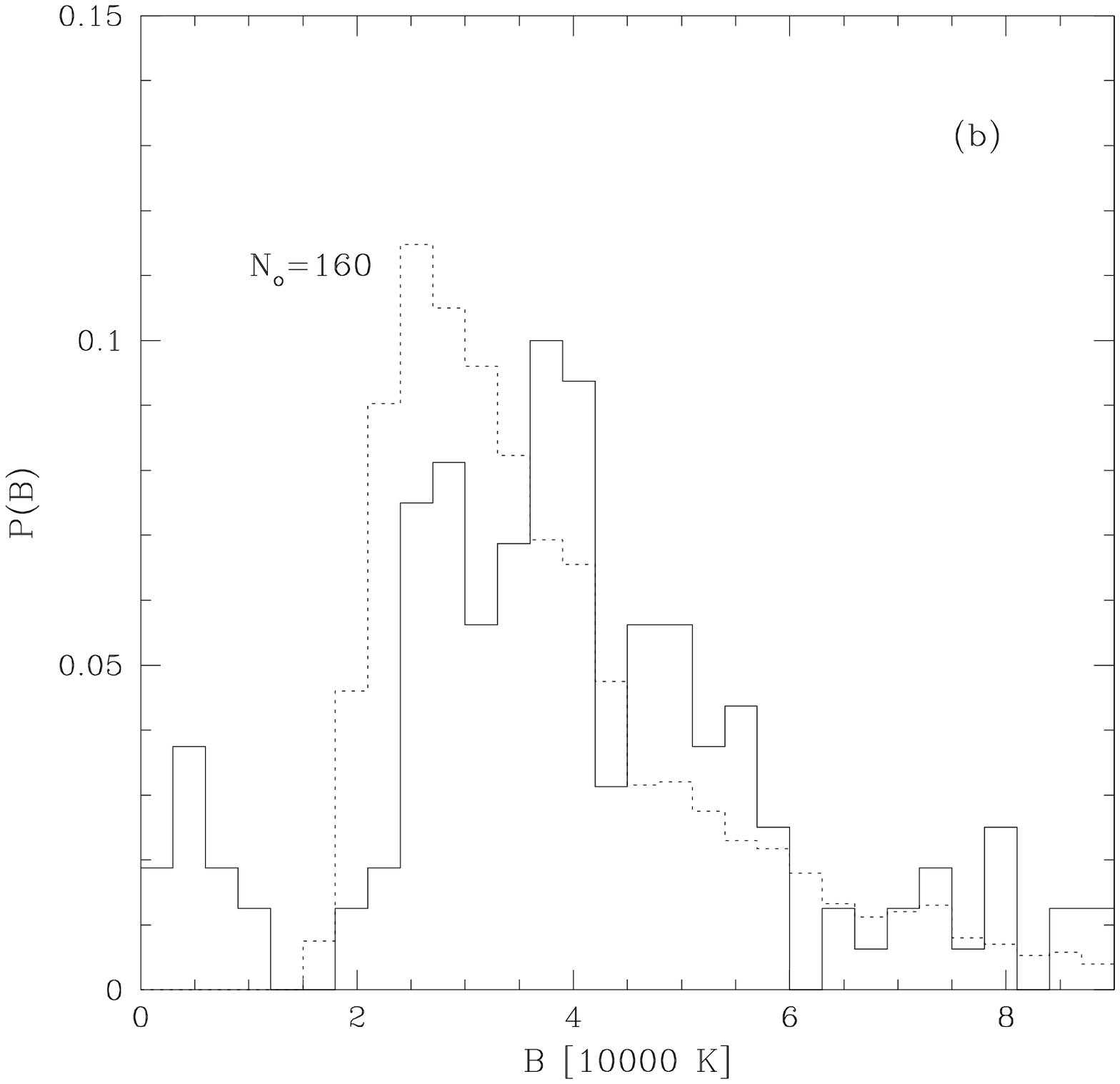}
\end{figure}
\begin{figure}
\plotone{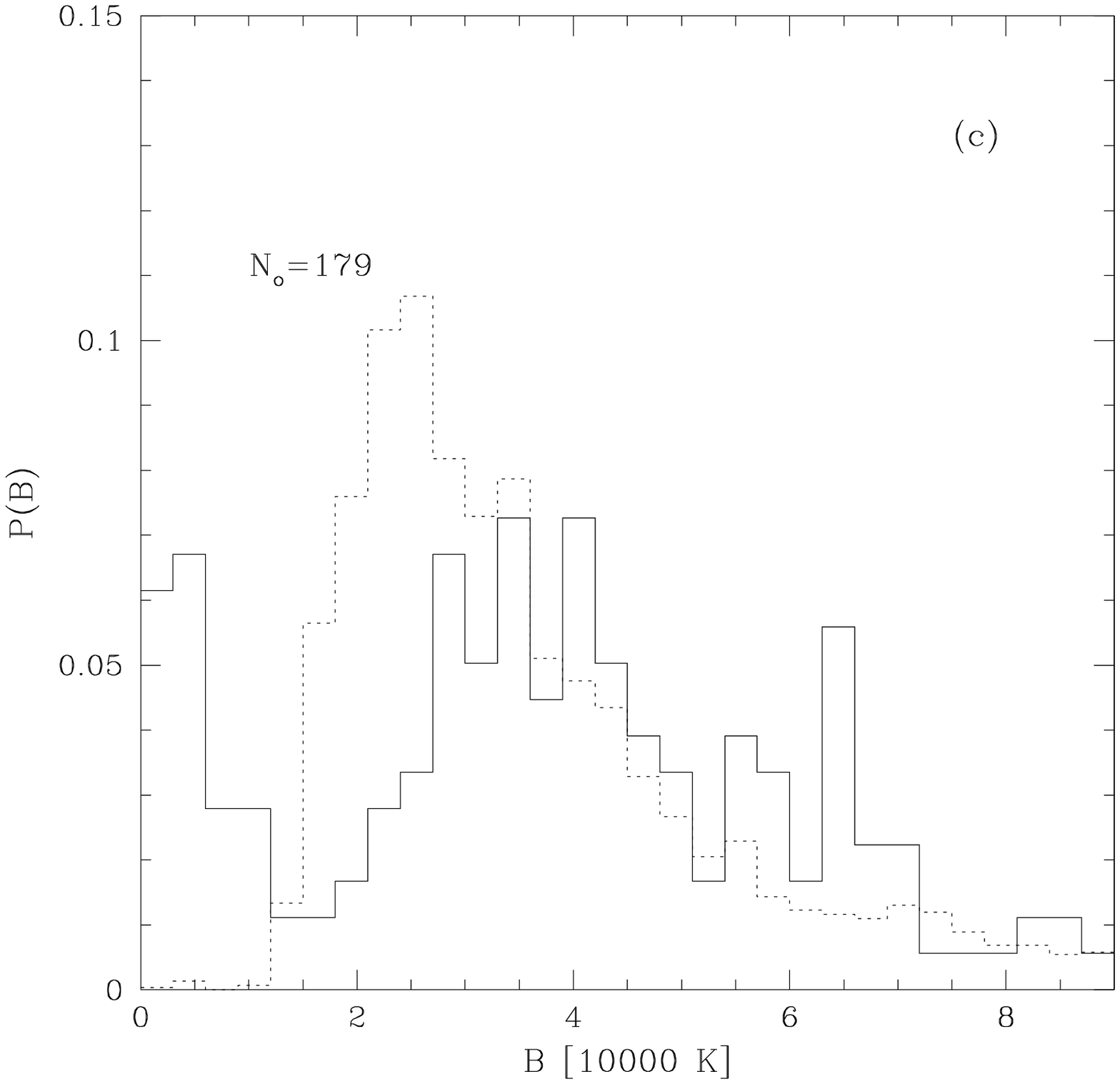}
\end{figure}
In Figure \ref{z3compsimobshist}(a,b,c) we compare the $B$ 
histograms of the observed lines, within the ranges of optical depth that 
we will use for the temperature measurements, to the $B$ histograms
of fitted lines from 
the simulation outputs with redshifts closest to the means of the 
observations, 
in the same optical depth ranges.  The
observed absorption lines obviously have higher temperatures than 
the simulated ones in all three cases.

It is interesting to compare directly $B_C(\tau_c)$ from the 
observations and from the simulation, 
before we determine $T=T_\star (\Db/\Delta_\star)^{\gmo}$ using the
more involved method described in \S 5.4.
\begin{figure}
\plotone{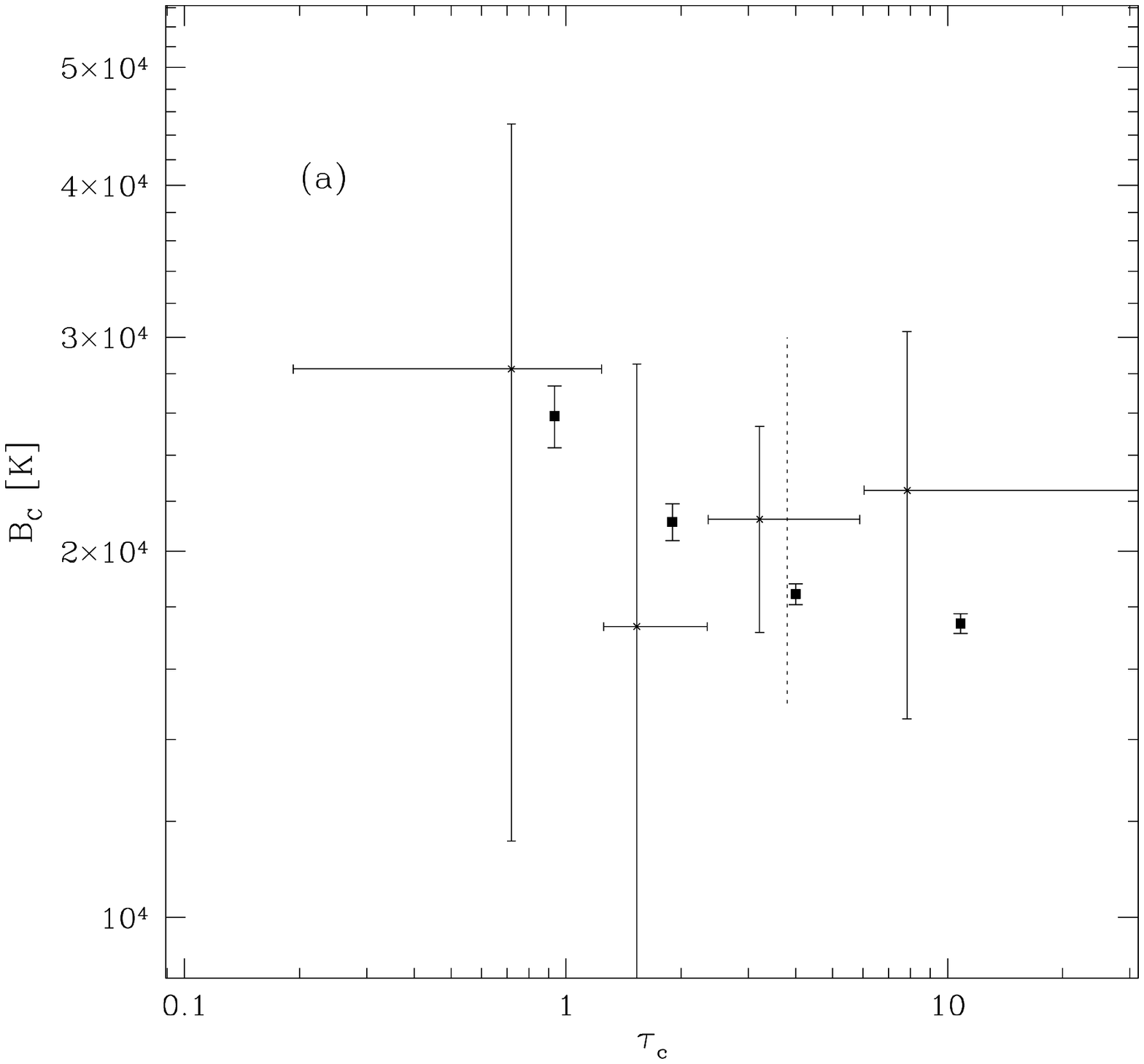}
\caption{(a, b, c) {\it Crosses with large error bars:} $B_C(\tau_c)$ for
observed lines at $\bar{z}=$(3.9, 3.0, 2.4).
{\it Squares with small error bars:} $B_C(\tau_c)$ from the simulation
at $z=$(4, 3, 2).  
Only lines in the range of $\tau_c$ between the vertical dotted lines
[or to the right of the single dotted line in (a)] are
used for our final temperature measurement.
}
\label{z3obsTest}
\end{figure}
\begin{figure}
\plotone{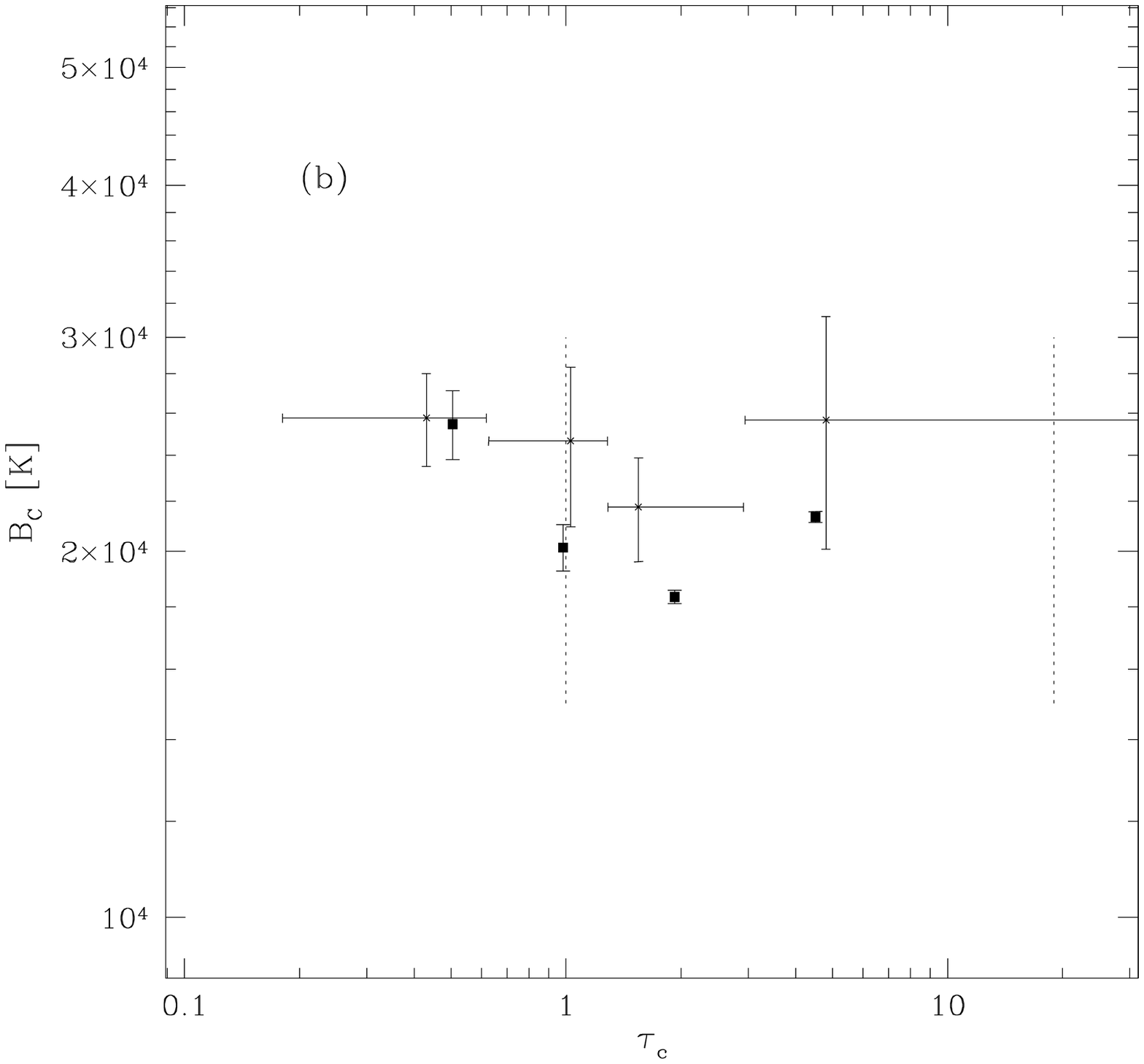}
\end{figure}
\begin{figure}
\plotone{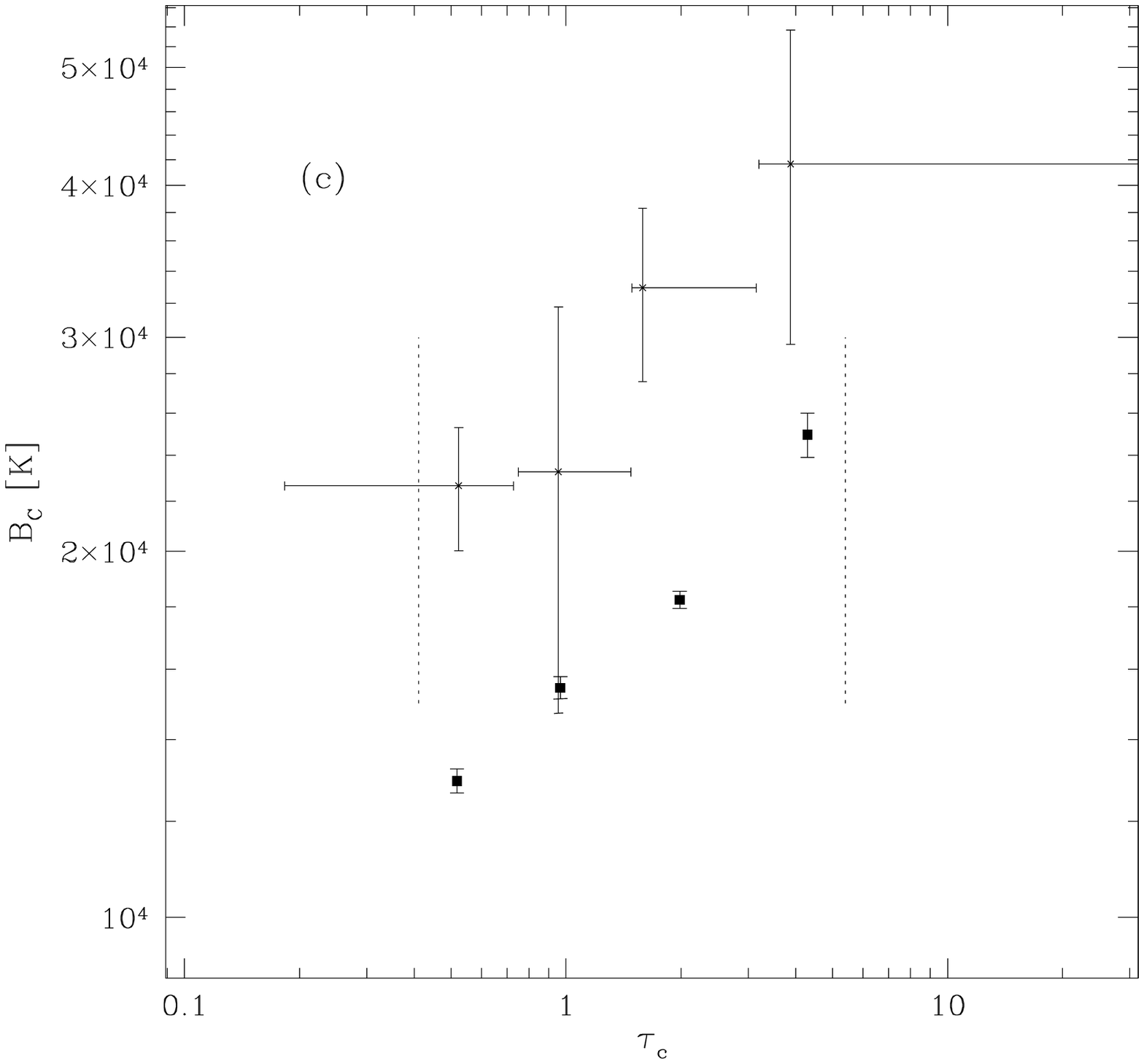}
\end{figure}
Figure \ref{z3obsTest}(a,b,c) shows the values and errors of $B_C$
measured from the observations and the simulation output that is nearest
in redshift, using lines over the optical depth bins indicated by the
horizontal error bars (the optical depth bins contain equal numbers of
observed lines). The vertical dotted lines indicate the optical depth
range that we use for the final temperature measurement
(for the $\bar{z}=3.9$ analysis, the upper limit on $\tau_c$ is outside
of the figure, and eliminates a negligible number of lines).
Recall (Figure \ref{Testerrz3}) that
we do not expect the points with lower $\tau_c$ to accurately
reflect the real temperature, except at the lowest redshift.  
The observed lines again appear
to be hotter than the simulated lines.  
These results are listed in Table \ref{bigtab}, along with the 
temperature offset $\Delta T$ used as a correction to the temperature
(see \S 5.4.1) and the estimated gas densities, $\Db$,
that we find once the \TDr in the simulation has been adjusted to 
match the observed one (determined below).
If more observed spectra were
available, this method of binning in optical depth would be preferable
to the method in \S 5.4 because it does not require the assumption of
a power-law \TDr.  Each bin in optical depth would be associated with the
density listed in Table \ref{bigtab}, and the value of $B_C$ would be
corrected by the listed $\Delta T$. 

We now determine $T_\star$ and $\gmo$ by the method described in 
\S5.4. In order to obtain the $\tau-\Db$ relation and the temperature
offsets $\Delta T$, we can use any of the three simulation outputs at
$z=2,3,4$ for any of the three redshift bins in which we have divided
the data (although we change the mean flux decrement of the simulated
spectra to the observed one at each redshift bin). The different
redshift outputs of the simulation are approximately equivalent to
assuming different models with a different amplitude of the power
spectrum (\mmr), so we can use the different outputs
to check that our measurement of the temperature is not greatly
sensitive to the model that is assumed. \mmr found that the
amplitude of the initial
density perturbations in our simulation needs to be reduced by about
15\% to agree with the observed power spectrum of the transmitted flux,
meaning that the simulation output at $z=4$ has fluctuations that
most closely match the observed ones at $\bar{z}=3$, and are slightly 
higher than the observed fluctuations at $\bar{z}=3.9$. 
The $\bar{z}=2.4$ observational bin
is closest in amplitude to the simulation output at $z=3$.
The most reliable temperature results should therefore be obtained by
using the $z=$(4, 4, 3) simulation outputs to analyze the
$\bar{z}=$(3.9, 3.0, 2.4) observations, but we shall also give results
for $\bar{z}=$(3.0, 2.4) analyzed using the $z=$(3, 2) simulation outputs. 

The result at $\bar{z}=3$, using the $z=4$ simulation output to 
predict the  $\tau-\Db$ relation and the necessary correction for 
non-thermal broadening, is:
\begin{equation}
T=[20200\pm 1300]\left(\frac{\Db}{1.37\pm 0.11}\right)^{[0.29\pm 0.30]} 
{\rm K}~,
\end{equation}
where the error bars on $T_\star$ and $\gmo$ are uncorrelated (which
defines $\Delta_\star$).  The error bar on 
$\Delta_\star$
reflects only the uncertainty
in the \mmr determination of the mean flux decrement, which
affects the relation between density and optical depth.  
This result implies $T_0=18400\pm 2100$K, where $T_0$ is the temperature
at the mean density. The pivot density, $\Delta_\star=1.37$, corresponds
to an optical depth $\tau_\star=1.83$ (from the relation obtained as
described in \S 5.4.2)
When we repeat the fitting using the $z=3$ simulation output we find
$T=[19600\pm 1500](\Db/[1.24 \pm 0.10])^{[0.33\pm 0.28]}$K, 
or $T_0=18300\pm 1800$K ($\tau_\star=1.74$).  
The difference between the two values for
$\Delta_\star$, 1.37 using the $z=4$ simulation, and 1.24 using the
$z=3$ simulation, are mostly a reflection of the different optical depth 
normalizing factors (i.e., rescalings of the baryon density or the 
strength of the ionizing background) 
needed to match the observed mean flux decrement.
The normalizing factor is smaller for the $z=4$ simulation,
giving a larger $\Delta_\star$ for the same optical depth, because 
the density fluctuations are of lower amplitude, leading to less
saturated absorption and more absorption in voids
(see \mmr for a more detailed
discussion of the optical depth normalizing factor). 

All the results obtained at the three redshift bins are listed in
Table \ref{plfittab}. The data analysis at $\bar{z}=2.4$ and
$\bar{z}=3.9$ is similar to the analysis at $\bar{z}=3$, except for the
differences that we mention below.

  The temperature results at $\bar{z}=2.4$ differ by $\sim 2000$ K when
the $z=2$ simulation output is used instead of the $z=3$ output, once
the measured values of $T_\star$ for the two are extrapolated to the
same density. Most of this difference results from the difference in
$\Delta_\star$, and the relatively high value of $\gmo$ that is obtained
at $\bar z = 2.4$. The difference in $\Delta_\star$ is caused by the
different amplitudes of density fluctuations in the two simulation
outputs, so we expect that the temperature derived using the $z=3$
simulation output, which has the correct amplitude of fluctuations,
is more reliable (see \mmr). 

  In order to see the evolution of the temperature with redshift, we
need to obtain the temperature at a fixed over-density at each redshift.
It is useful to obtain the temperature at the mean density, $T_0$, to
compare our results to other work. However, the values of $\Delta_\star$
are close to $\Db=1.4$ at all three redshift bins, and we can therefore
have a more robust result for the temperature evolution if we examine
the temperature at $\Db=1.4$, which we denote as $T_{1.4}$ in Table
\ref{plfittab}. 
 
  We note here that, because of the small range of optical depth used at
$\bar{z}=3.9$, at this redshift we were forced to smooth
the $B''$ histograms with a Gaussian filter of width $\sigma_B=7000$K, instead
of our standard $\sigma_B=5000$K, in order to avoid problems with multiple, 
approximately equivalent, maxima of 
equation (\ref{difeq}) as $\gmo$ is varied. 

\begin{figure}
\plotone{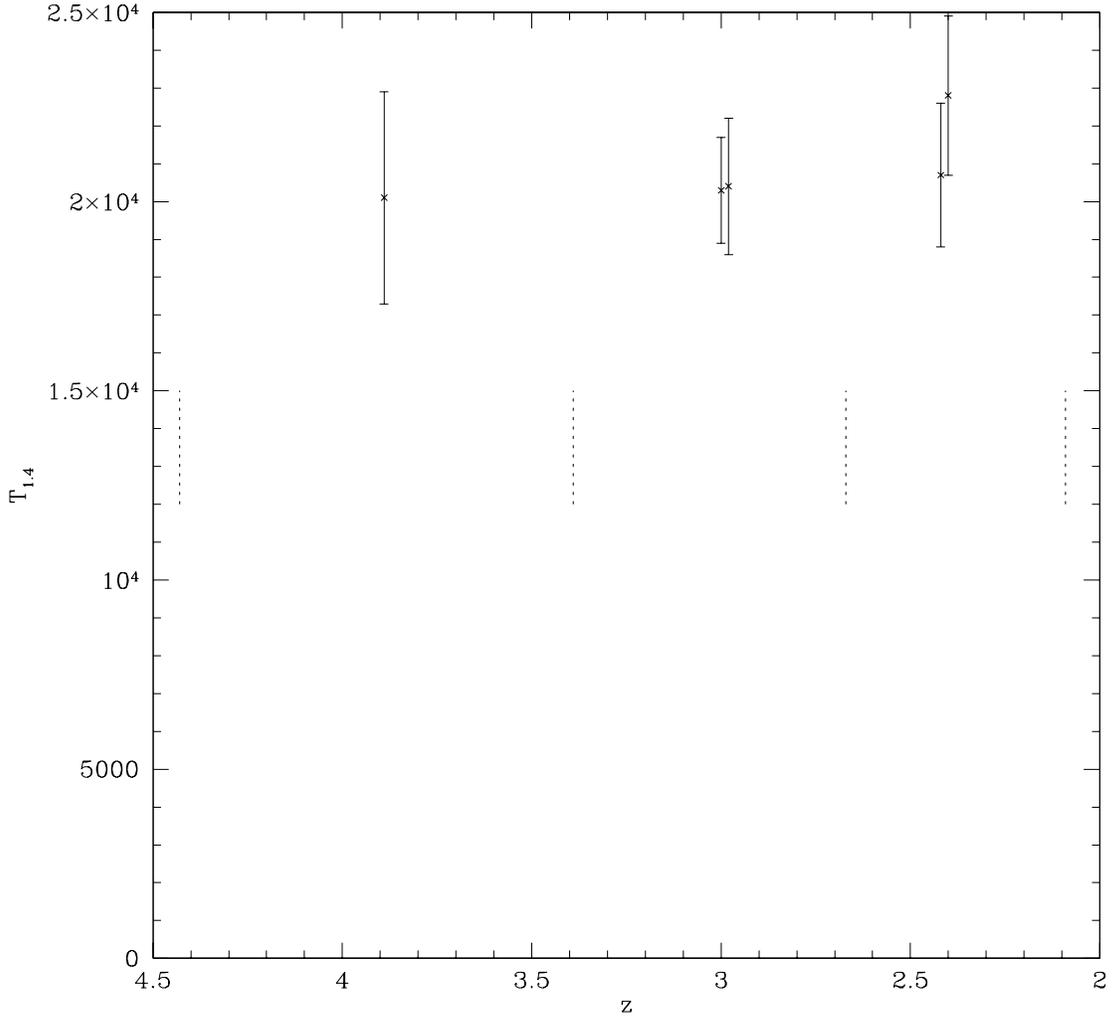}
\caption{The observed temperature at $\Db=1.4$.  The
two points at $\bar{z}=2.4$ and $\bar{z}=3.0$ (offset slightly
to distinguish them) show the result of using two
different simulation outputs to analyze each of these redshift bins.
The left and right points at $\bar{z}=2.4$ are for the $z=3$ and 
$z=2$ outputs, respectively.  The left and right points at 
$\bar{z}=3.0$ are for the $z=4$ and $z=3$ simulation outputs.
The vertical dotted lines show the boundaries of the redshift bins.
}
\label{Tresultsfig}
\end{figure}
\begin{figure}
\plotone{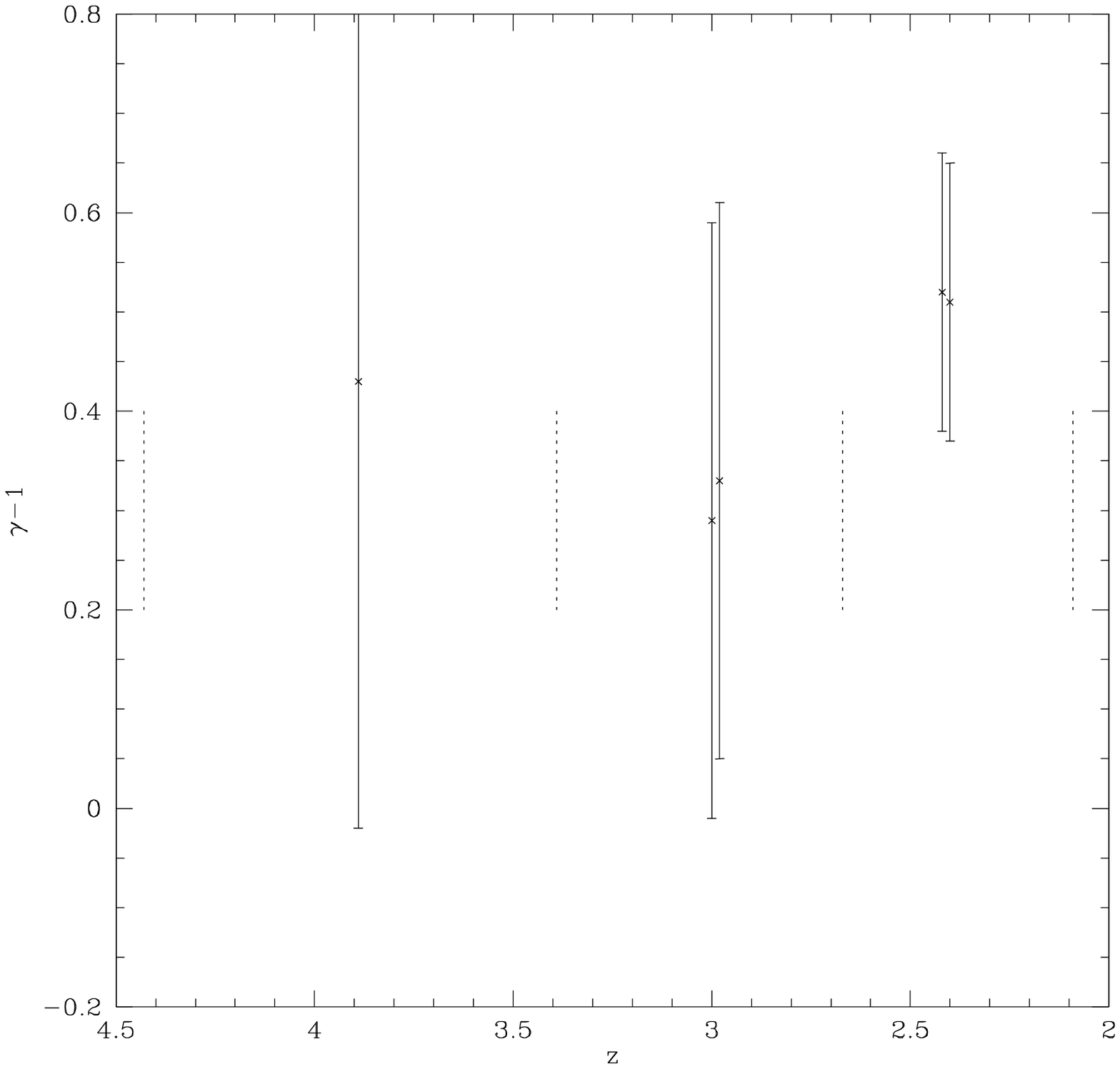}
\caption{Similar to Figure \ref{Tresultsfig} except here we plot
the results for $\gmo$.
}
\label{gmoresultsfig}
\end{figure}
The primary results of this paper, the measurements of $T_{1.4}$
and $\gmo$, are summarized in Figures
\ref{Tresultsfig} and \ref{gmoresultsfig}, respectively. 
These show two important conclusions: first, the temperatures are
higher than the value expected if photoionization heating in
equilibrium is the only heating source. Second, we find no evidence
for a rapid change of the temperature with redshift.

\section{DISCUSSION}

  This paper presents a measurement of the temperature-density relation
of the intergalactic gas in the redshift range $2.4 < z < 4$. The new 
method we have developed to perform this measurement is based on the
same general idea as the previous work by 
\citet{stl99}, \citet{rgs99}, and \citet{bm99}: provided that there is
a tight relation between the temperature and density of the gas,
absorption lines of similar central optical depth should have little
dispersion in their thermal broadening, and the varying line widths
should correspond to variable amounts of hydrodynamic broadening.
Occasionally, some absorption lines will be subject to only a small
degree of hydrodynamic broadening; this will typically happen when most
of the atomic hydrogen occurs near a velocity caustic along the line of
sight. We therefore expect the histogram of line widths to show a rapid
increase near the value of the Doppler parameter corresponding to the
gas temperature. In the absence of noise, every line should be
wider than the thermal broadening width, so at least an upper limit to
the temperature can be obtained unambiguously.

  The tests we have performed using a numerical simulation of the \lya
forest, based on a CDM model that successfully reproduces the
observations of large-scale structure at present, confirm this general
idea. However, they show that this method to recover the gas
temperature works efficiently only over a limited range of line optical
depths, which corresponds approximately to a range of gas over-density
$1 \lesssim \Delta_g \lesssim 3$. At lower densities, the gas is
generally in Hubble expansion and this effect dominates the contribution
to the line widths in essentially all the lines. The minimum line widths
can therefore only provide an upper limit to the temperature of this
low-density gas. Of course, the simulation can in principle be used to
correct for the effect of line broadening due to Hubble expansion, and
to obtain the temperature by subtracting the hydrodynamic contribution
to the minimum line widths. However, the results can then strongly
depend on the assumed model and the numerical resolution of the
simulation, especially as the thermal broadening becomes a small effect
compared to the expansion.

  At very high densities, the increasing dispersion of the temperature
at a given gas density can result in a large difference between the
typical gas temperature and the ``lower cutoff'' in the line width
histogram. This implies again that the recovery of the median gas
temperature from the distribution of line widths is highly sensitive to
the model assumptions that affect the temperature dispersion and the
turbulent motions in the gas.

  There are two main differences between the method we use to measure
the gas temperature, and that used by previous authors. First, our
line detection algorithm avoids the necessity of the Voigt-profile
fitting method to ``deblend'' lines, by simply throwing out any
``absorption lines'' that do not correspond to a clearly identified
minimum in the transmitted flux, or that do not have a large enough
region around that minimum that is adequately fitted by a simple
Gaussian in optical depth. In the method we use here, every line width
is essentially a measurement of the second derivative around a minimum
in the flux. This is one important reason why the total number of lines
we identify is much lower than \citet{str00}, even though we use nearly
the same data set. The second difference is that we restrict the
absorption lines we use to lie within the range of central optical
depth where the correction that needs to be applied to the temperature
measured from the histogram of line widths (as described in \S 5.4.1)
is small.

  These differences explain our substantially increased error bars in
measuring $T_0$ and $\gamma$, relative to those of \citet{str00}.
However, we believe our error bars are more reliable and
model-independent, for the reasons we have discussed. We also note here
that, even though we cannot rule out a substantial change of the
temperature we have measured depending on the numerical simulation of
the \lya forest that is used for comparing to the observational data,
this possibility appears unlikely for several reasons, in addition to
the arguments explained before about the small size of the correction
$\Delta T$ that needs to be applied to the line width cutoff $B_c$ to
obtain the gas temperature (see \S 5.4.1). The gas temperature of the
simulation we use is lower than that observed only by a small amount: in
the simulation, the temperature at the mean density is $T_0$=(14000,
15800, 12800)K at $z=$(4, 3, 2), while our measurement from the
observational data is $T_0=(17400\pm 3900$, $18400\pm 2100$, $17400\pm
1900$)K for $\bar{z}=$(3.9, 3.0, 2.4). The amplitude of the flux power
spectrum of the simulation is also very close to the observationally
determined one (\mmr). We have also shown that we obtain nearly
identical results for the temperature measurement when using different
simulation outputs to analyze the same observations (see Figures
\ref{Tresultsfig} and \ref{gmoresultsfig}) meaning that the dependence
on the amplitude of the power spectrum is weak. Our result for the gas
temperature might also be affected by the limited resolution of the
simulation we use (with a comoving cell size of $35$ Kpc). We have not
yet performed a convergence test for the effects of resolution on the
\lya forest; however, \citet{str00} find that a mean particle spacing of
$\sim 45$ Kpc in their SPH simulations is sufficient for convergence.

  We now compare our results for the evolution of the gas temperature
with previous measurements. \citet{rgs99} find a temperature at the mean
density (from their Figure 12b) $T_0 \simeq$($18600_{-6900}^{+10900}$, 
$23400_{-5200}^{+10400}$, $17000_{-9600}^{+22800}$)K, at $z\simeq$
(3.6, 2.75, 1.9). Considering their large statistical error bars,
our results appear to agree well with theirs, although the
true temperature at $z\simeq 2.75$ must be at the low end of their error bar.
\citet{str00} give their temperature results as 16 separate points, in their
Figure 5.  We read off their points in each of our redshift bins and create 
error weighted averages for comparison with our results, 
finding $T_0\sim$($12200\pm 1700$, $17300\pm 1400$, $14000\pm 1300$) at
$z\sim$(3.75, 3.2, 2.4). In order to compare these with our temperatures,
which have lower errors at $\Db=1.4$, we extrapolate
the \citet{str00} results to $\Db=1.4$ using their
measured value of $\gmo$, obtaining $T_{1.4}\sim 13600\pm 2000$K at
$\bar{z}\sim 3.75$.  Our result at $\bar{z}=3.9$, $T_{1.4}=20100\pm 2800$K,
is higher than theirs by $\sim 1.9 \sigma$. Because of this, we
do not find evidence for the increase of the temperature with
time between $z=4$ and $z=3$ that \citet{str00} reported. Our results are
consistent with a constant temperature.

  Actually, the data set analyzed by \citet{str00} is almost identical
to the one we analyze in this paper, with 7 of the 10 quasars used in
the two papers being identical.  However, our methods of analysis are
very different, so that even the statistical error bars from the two
analysis are largely independent. The main difference, as mentioned
before, is in the number of lines that are identified. For example, 
we use 98 lines to measure the temperature in our $z>3.4$ bin, while 
\citet{str00} use about 550 lines in a comparable bin, even though
$\sim$ 80\% of the data that they use is identical to ours.

\subsection{HeII Reionization as a Heating Mechanism}

  Is the value of the temperature we have measured in agreement with
the known sources of heating and cooling in the intergalactic gas?
The evolution of the temperature is determined by the equation:
\begin{equation}
{ d \log T \over H\, dt } = -2 \left( 1 - {1\over 3}
 {d \log \Delta \over H\, dt} \right)
+ {2 \over 3 k H T}\, (L_{He} + L_H - L_{CMB} - L_R - L_{ff} - L_a) ~,
\label{tbal}
\end{equation}
where the cooling and heating rates per particle are denoted as
follows: $L_{He}$ is the heating by \heii photoionization, $L_H$ is the
heating by \hi photoionization, $L_{CMB}$ is the cooling off the
microwave background, $L_R$ is the cooling by recombination, $L_{ff}$ is
the cooling by free-free emission, and $L_a$ is the cooling due to line
excitation and collisional ionization. We have separated the terms for
\heii and \hi photoionization because \heii plays a dominant role for
heating, but the other cooling terms include both hydrogen and helium.

  To see what the important terms for the thermal balance of the IGM are,
we now evaluate all the heating and cooling terms at the conditions
where we have measured the temperature most accurately: a temperature
$T=20000$ K at a gas density $\Delta=1.4$, at $z=3$. It is convenient
to define the quantities $T' \equiv 2/(3kHT) L$, for each subscript
corresponding to every heating and cooling term. We assume the model
$\Omega_b h^2 = 0.019$, $\Omega_0=0.3$, $\Lambda_0=0.7$, and
$H_0=65 \kms\mpc^{-1}$. We evaluate first the
total cooling: the dominant term is adiabatic cooling, which is equal to
$-2$ on the right-hand-side of equation (\ref{tbal}) if we assume
a rate of expansion equal to the Hubble rate
(i.e., a constant $\Delta$). This assumption is of course not exact;
every gas element expands at a different rate, causing a dispersion in
the temperature-density relation. However, at a density $\Delta=1.4$, the
average rate of expansion is in fact not very different from the Hubble
rate. In addition, when we consider the evolution of the temperature at
a fixed $\Delta$, the $\partial T / \partial \Delta$ term in the total
temperature derivative of the left-hand-side of equation (\ref{tbal})
should partly compensate for the effect of expansion (canceling it
exactly if $\gmo = 2/3$).

  The next most important contribution is cooling off the microwave
background, which is given by
$T'_{CMB} = (8 \sigma_T a T_{CMB}^4)/(3 H m_e c) (n_e/n) = 0.58$ (where
$\sigma_T$ is the Thompson cross section, $m_e$ the electron mass, $n_e$
the electron density, and $n$ the total particle density).
Notice that this term becomes more important at higher redshift, growing
as $(1+z)^{5/2}$. We compute the other cooling rates using the formulae
given in \citet{b81}. Recombination yields $T'_R = 0.22$, and
free-free emission $T'_{ff} = 0.08$. The atomic processes of line
cooling and collisional ionization are completely negligible at this
low density, for the \hi photoionization rate $\Gamma\sim 10^{-12} \,
{\rm s}^{-1}$ that is obtained from the observed abundance of quasars.
The total cooling rate is therefore $T'=2.88$. If this
temperature is being kept roughly constant, as indicated by the
measurement we have presented here, then the heating terms should
approximately balance the total cooling.

  To evaluate the heating rate, we first assume ionization equilibrium;
we will discuss later how the heating from \heii can be
increased if the \heii reionization is still in progress. The heating
term due to ionization can then be expressed in terms of the
recombination rate: $L_H= <E_H> (\alpha_H n_e) (n_H/n)$, and the
analogous expression for \heii, where $\alpha_H$ and $n_H$ are the
recombination coefficient and the number
density of hydrogen, and $E_H$ is the mean energy of the absorbed
photons minus the ionization potential. The mean energy $E_H$ depends
on the spectrum of the ionizing background, and we evaluate it as
follows: assuming a background intensity per unit frequency
$J_{\nu} \propto \nu^{-\beta}$ from the ionization edge at $\nu_0$ to some
maximum frequency $\nu_m=q_m \nu_0$, and approximating also the
photoionization cross section as $\sigma(\nu) \propto \nu^{-3}$, we find
\begin{equation}
<E_H> = I_H\, \left[ { 1-q_m^{-\beta-2} \over 1-q_m^{-\beta-3} } \,
{ \beta + 3 \over \beta + 2 } \, -1 \right] ~,
\end{equation}
where $I_H$ is the ionization potential.
We use $\beta=0$ and $q_m=4$ for both \hi and \heii, which adequately
approximates the shape of the spectrum found in numerical calculations
when the emitted spectrum is a quasar power-law (with $\beta=1.5$), and
the effect of absorption by Lyman limit systems is taken into account
\citep{mo90,hm96}. This yields
$<E_H>=0.43 I_H$ and $<E_{He}> = 0.43 I_{He}$. We then
obtain: $T'_H=0.67$, and $T'_{He}=1.11$. The total heating therefore falls
short to compensate for cooling by a factor $\sim 1.6$.

  An obvious way to increase the heating rate is to assume that the
\heii reionization is not yet complete; in other words, that there are
patches of low-density gas in the IGM where all the helium is in the
form of \heii. As discussed in \citet{mr94}, there are two reasons why
the heating rate is higher during the reionization, relative to the case
of photoionization equilibrium. The first is that the ionization rate
needs to be higher simply because every \heii ion needs to be ionized
once during the course of the \heii reionization. The second is that
{\it all} the hard photons will now be absorbed by a random \heii ion in
the IGM, up to the frequency $\nu_m=q_m \nu_0$ where the mean-free-path
through the \heii IGM reaches the horizon length. For the baryon density
we use and at $z=3$, and assuming also that about 50\% of all the helium
is in the form of \heii in the diffuse IGM (and not in dense clouds
having a small covering factor over the Hubble length), this maximum
frequency is given by $q_m=13$, i.e., a frequency 13 times higher than
the \heii ionization edge. 
The mean energy of the absorbed photons is therefore equal to the mean
energy of the emitted photons up to this maximum frequency,
{\it without} weighting them with the photoionization cross section:
\begin{equation}
<E_{He,r}> = I_{He}\, \left[ { 1-q_m^{-\beta'+1} \over 1-q_m^{-\beta'} } \,
{ \beta' \over \beta' - 1 } \, -1 \right] ~,
\end{equation}
where the subscript $r$ in $<E_{He,r}>$ indicates the mean energy per
absorption during reionization, and the emitted spectrum from sources is 
$J_{\nu} \propto \nu^{-\beta'}$. For $q_m=13$ and $\beta'=1.5$, we
obtain $ <E_{He,r}> = 0.71 I_{He}$. Since the recombination rate for
\heii at the mean density is equal to $1.6$ times the Hubble rate
$H^{-1}$ at $z=3$, and if the reionization is occurring over $\sim$
a Hubble time near $z\sim 3$, it is reasonable to expect that the
additional heating rate due to \heii reionization is comparable to the
heating rate due to balancing recombinations of \heii. We therefore
conclude that the heating from \heii reionization can reasonably
account for the IGM temperature we have determined here.

\subsection{Usefulness for Measuring the Baryon Density from the
Lyman Alpha Forest}

One of the applications that the development of the new theory of the
Ly$\alpha$ forest based on structure formation has had is to provide
a measurement of the baryon density through its effect on the mean
transmitted flux. 
For a fixed distribution of temperature, over-density, and peculiar
velocities in the IGM, the \lya optical depth at any point in the
spectrum is proportional to
$\left(\Omega_B h^2 \right)^2 H(z)^{-1} \Gamma_{-12}^{-1}$,
where $\Gamma=10^{-12} \, \Gamma_{-12}\, {\rm s}^{-1}$ is the
photoionization rate due to the cosmic ionizing background.
To be specific, we define the parameter
\begin{equation}
\omega_B \equiv \Omega_B h^2 
\left[\left(\frac{\Omega_0 h^2}{0.3 \times 0.65^2}\right)^{1/2}
\Gamma_{-12}\right]^{-1/2}~,
\end{equation} 
[where we have used $H(z)\simeq H_0 (1+z)^{3/2} \Omega_0^{1/2}$, which
is highly accurate at the relevant redshifts and in a flat universe].
As discussed by \citet{hkw96}, \citet{mco96}, \citet{rms97}, 
\citet{wmh97}, and \mmr, a measurement of $\omega_B$ 
can be translated into a lower bound on $\Omega_b h^2$ by using the
contribution to the ionizing background from known quasars as a lower
bound on $\Gamma_{-12}$.
This lower limit is on the high side of the
range of $\Omega_b$ that is allowed by primordial nucleosynthesis:
$\Omega_b h^2 \gtrsim 0.02$ \citep[][; \mmr]{rms97}.

  One of the main model uncertainties in deriving the relationship
between the parameter $\omega_B$ and the predicted mean transmitted
flux is the mean IGM temperature. A higher temperature implies a lower
recombination coefficient, and therefore a lower neutral hydrogen
density. This implies that in order to reproduce a given observed
mean transmitted flux, the mean density $\Omega_B$ needs to be
increased further to compensate the reduced recombination
coefficient. The measurements of the IGM temperature reported here
and in \citet{str00}, and \citet{rgs99}, all coincide
in finding temperatures that are high compared to what is expected if
the IGM is heated by photoionization and is in ionization equilibrium.
As we have discussed, these higher temperatures can probably be
understood as a result of the \heii reionization.
independently of its cause, the higher temperature implies an even
higher value of $\omega_B$ than was obtained previously, which we can easily
determine by modifying the temperature in the simulation to match the
observations, as described earlier in \S 5.3. We find that, when the
temperature in the simulation is increased to match the observed one,
the derived value of $\omega_B$ is increased slightly to
$\omega_B=$($0.0336\pm 0.0020$, $0.0288\pm 0.0023$, 
$0.0248 \pm 0.0017$) at
$\bar{z}=$(3.9, 3.0, 2.4), from the previous values 
(0.0329, 0.0274, 0.0245) when the
temperatures in the simulations are not modified (\mmr).
The errors are derived from the observational errors in the
determination of the mean flux decrement from \mmr. 

  Changing the temperature of the simulation affects the value of 
$\omega_B$
not only by modifying the recombination coefficient, but by increasing
the amount of thermal broadening, which can spread the absorption in
saturated regions to the outskirts of absorption lines, increasing the
mean absorption for a given $\omega_B$. We find that the effect of
thermal broadening is less important than the effect of the reduced
recombination coefficient. As an example,
if we replace $\alpha(T)\rightarrow
\alpha(T+3000K)$ in every pixel in the simulation,
the inferred $\omega_B$ increases by 0.0017 (where $\alpha(T)$ is the 
recombination coefficient), while the replacement 
$\sigma_b(T)\rightarrow \sigma_b(T+3000K)$ changes $\omega_B$ 
by -0.00046 
(where $\sigma_b(T)$ is the dispersion of the Gaussian thermal
broadening).
  Dynamical effects caused by the increased pressure of the gas would
probably go in the same direction as the thermal broadening, since they
would tend to spread the gas in absorption systems over wider regions.
However, it seems unlikely that any such dynamical effects (which can
only be investigated by running the same simulation with different
temperatures) can be more important than the thermal broadening effect.
The examples shown in \citet{tsh99} (see their Figure 6) appear to 
confirm that the dynamical effects of increased
pressure are not more important than the increased thermal broadening
when the gas temperature is raised.

 With the statistical error bars we have obtained on the \TDr, we can
place more conservative lower bounds on $\omega_B$ than the ones
obtained in \mmr. The lowest allowed value of $\omega_B$ needed to
account for a given observed mean transmitted flux is obtained when
$T_\star$ is minimum and $\gmo$ is maximum in equation (1), because
that yields the minimum temperature for the low-density gas that
determines the optical depth in unsaturated regions of the \lya
spectrum. We set $T_\star$ equal to the measured value minus
twice the statistical error bar given in Table \ref{plfittab}, and
$\gmo=0.6$, which is the value valid when the IGM has been in
photoionization equilibrium for a long time
\citep{hg97}. Any uniform heating of the IGM, such as that
caused by reionization, should give rise to a lower $\gmo$,
although shock-heating can increase $\gmo$ above $0.6$, the simulations
show that this happens only at high enough gas densities that the
\lya absorption is already saturated. The error in our observational
determination of $\gmo$ is too large to give us a better constraint
than $\gamma < 0.6$ (see Table \ref{plfittab}).

  The results of this exercise are $\omega_B>$(0.0270, 0.0192, 0.0209 )
for $\bar{z}=$(3.9, 3.0, 2.4), at 95\% confidence, including the
error from the mean flux decrement and the temperature measurements
(added in quadrature).
Using the lower bound obtained in \citet{rms97} of $\Gamma{-12} > 0.7$
in the range $2 < z < 3$,
obtained by counting only radiation from the observed quasars, and not
including the power-law extrapolation of the quasar luminosity function
that has been observed only at redshifts $z<2$, we obtain
$\Omega_B h^2 > 0.017$. This result is still consistent with the
determinations of the deuterium abundance \citep{bt98}.
However, if the quasar luminosity function extends to low luminosities
with a similar power-law slope as observed at $z<2$, or if emission from
galaxies increases significantly the intensity of the ionizing
background, then the higher baryon density implied would come into
conflict with the primordial nucleosynthesis predictions and the
observed deuterium abundance.

In summary, we have reached the following conclusions:\\
1.  The temperature of the IGM is $\sim 20000\pm 2000$K at density 1.4 
  times the mean, independent of redshift, although an increase of 
  $\sim 3500$K from $z=3.9$ to $z=3.0$ cannot be ruled out. \\
2.  The high temperature cannot be explained by heating in ionization
  equilibrium, and probably indicates on-going \heii reionization. \\
3.  The contribution of temperature uncertainty to the uncertainty in 
  the  baryon density required by 
  the observed mean flux decrement in the \lya forest is now well
  constrained. 
  
We thank Adam Steed and David Weinberg for helpful comments on the
manuscript.  

\appendix
\section{THE PROFILE FITTER}

This algorithm has three input parameters that control how 
the fitting proceeds:  
$E_d$ sets the amount that the flux must increase from the center point
to the edges of the window before a fit will be attempted, 
$W_{min}$ sets the minimum size of the window 
within which a fit is performed,
and $P_0$ sets the quality of fit that will be accepted.
In this paper we set $W_{min}=2$.

Three more input parameters effect the speed of the code but are not 
important to the results:
$E_s$ controls the degree of symmetry around a central pixel
that is required for a fit to be attempted,
$E_c$ sets the level of flux decrease, from the center pixel to the 
window edges, at which a point will be eliminated from consideration
for fitting, and
$W_{max}$ sets the maximum allowed window size.  These parameters are
set to values large enough that they do not actually eliminate any
profiles that would otherwise be accepted.

Before we describe the algorithm in detail, a few more terms must be 
introduced:
We are going to fit pieces of the spectrum that 
have center point $P$ and extend $\pm W$ pixels to either side of $P$.
The width of the fitting window, $W$, will be adjustable but 
constrained to $W_{min}\leq W \leq W_{max}$.
The transmitted flux at a point $P$ is $F(P)$.  The error in the sum 
or difference of the flux at two points $P_1$ and $P_2$ is 
$\sigma(P_1,P_2)=[\sigma(P_1)^2+\sigma(P_2)^2]^{1/2}$, where 
$\sigma(P)$ is the observational error in the flux at point $P$.
The minimum acceptable probability for $\chi^2$ is $P_0$ (we need
to define $P_0$ by the probability because there will be varying 
numbers of degrees of freedom in the fits).

For a given spectrum
the algorithm that we use is the following
(the reader should keep in mind that, except for the added 
complication of setting the window position and width, this procedure
just fits a single Voigt profile to each absorption maximum  by
$\chi^2$ minimization):\\ \\
1.   Scan along the spectrum pixel by pixel searching for places where 
$|F(P-W_{min})-F(P+W_{min})|<E_s~\sigma(P+W_{min},P-W_{min})$.  
Also require that $F(P)-F(P\pm W_{min})<
E_c~\sigma(P,P\pm W_{min})$, 
where the flux at the $\pm W_{min}$ points is averaged.
These places are candidates for a symmetric, non-concave profile.\\  
2.  If there is a significant increase in flux at the edges of the
window, so that $F(P)-F(P\pm W_{min})>
E_d~\sigma(P,P\pm W_{min})$, go ahead and fit Equation (\ref{profeqn})
to the absorption.  If
there isn't a significant increase try to expand the window.\\
3.  To expand the window require that symmetry is maintained when
$W$ is increased,
i.e., $|F(P-W)-F(P+W)|<E_s~\sigma(P+W,P-W)$.  If
the window can be expanded return to step 2 to check if 
a fit can be done with the enlarged window, i.e.,
if $F(P)-F(P\pm W)>
E_d~\sigma(P,P\pm W)$.\\
4.  If the region can't be fit, but also can't be expanded, 
eliminate the candidate point.  Also eliminate the point if 
the window size has been increased to $W_{max}$ without meeting the
requirement for fitting.\\
5.  Set initial parameters for the fit using $F(P)$ to set $\tau_c$ and
    $[F(P+W)+F(P-W)]/2$ to set $\sigma_b$.  Set  
    $v_c=0$.  If $F(P)<0$ set 
    $\tau_c=10$.  \\
6.  Minimize $\chi^2$ using the flux values and their error bars in
    the range of points between $P+W$ and $P-W$.  Require that 
    $|v_c|<0.5$ pixels (outside this range is covered by other 
    candidate points).\\
7.  Eliminate candidate if $P(>\chi^2,\nu)<P_0$.\\
8.  If $P$ falls within $W$ of a previously accepted candidate, 
    eliminate
    the candidate with a smaller value of $P(>\chi^2,\nu)$.  This does
    not eliminate any independent profiles because a single Gaussian
    would not fit if the window contained multiple lines.

\begin{deluxetable}{lcccc}
\tablecolumns{5}
\tablecaption{Power-law fits to the \TDr. \label{basesimTdbtab}}
\tablehead{
\colhead{$z$} & \colhead{$T_0$} & \colhead{$\gmo$ (for $T$)} & 
\colhead{$\tilde{T}_0$} & \colhead{$\gmo$ (for $\tilde{T}$)} \\ 
&(K)&&(K)&}
\startdata
4 & 14024 & 0.21 & 13873 & 0.22 \\
3 & 15843 & 0.30 & 15584 & 0.30 \\
2 & 12764 & 0.57 & 12954 & 0.54 \\
\enddata
\tablecomments{$\tilde{T}$ is the optical-depth-weighted average 
temperature
at points in spectra (fitted vs. the density similarly averaged).}
\end{deluxetable} 
\begin{deluxetable}{lccc}
\tablecaption{
\label{Qchanges}}
\tablehead{
\colhead{Setting} & \colhead{$N_g$} & \colhead{$N_b$} & \colhead{$Q$}}
\startdata
$E_d=25$        & 198 & 3   & 13.8 \\
$E_d=20$        & 236 & 11  & 14.3 \\
$E_d=17$        & 248 & 19  & 14.1 \\
$E_c=15$        & 264 & 22  & 14.3 \\
$E_c=14$        & 272 & 28  & 14.1 \\
$E_c=13$        & 271 & 34  & 13.6 \\
$E_d=12$        & 281 & 40  & 13.5 \\
$E_d=11$        & 285 & 56  & 12.4 \\
$E_d=8$         & 364 & 95  & 12.6 \\
$E_d=7$         & 386 & 137 & 10.9 \\
$E_d=5$         & 474 & 263 &  7.8 \\
$E_d=12$ (0.001)& 284 & 41  & 13.5 \\
$E_d=12$ (0.1)  & 243 & 35  & 12.5 \\
$E_d=12$ (nn)   & 236 & 1   & 15.3 \\
$E_d=12$ (nc)   & 262 & 38  & 12.9 \\
$E_d=12$ (nr)   & 286 & 39  & 13.7 \\
$E_d=12$ (nr)   & 306 & 30  & 15.1 \\
$E_d=12$ (nr)   & 308 & 35  & 14.7 \\
\enddata
\tablecomments{ 
The quality measure
$Q=(N_g-N_b)(N_g+N_b)^{-1/2}$, where $N_g$ is the number of 
fitted features
satisfying $0~{\rm K} < B-T < 3000$ K and $N_b$ is
the number satisfying  
$-3000~{\rm K} < B-T < 0$ K.
Entries labeled (0.1) and (0.001) have 
$P(>\chi^2)>$0.1 and 0.001, respectively
[the rest have $P(>\chi^2)>0.01$].  The label (nn) means no noise, 
(nc) means no continuum fitting approximation,  
and (nr) means a new set of random numbers was used for the added 
noise in each example.
}
\end{deluxetable}  
\begin{deluxetable}{lccccccccccc}
\tablecolumns{12}
\tablecaption{Basic statistics of the observational data in each 
redshift bin. 
\label{basicobsstats}}
\tablehead{
\colhead{$z_{min}$} & \colhead{$z_{max}$} & \colhead{$\bar{z}$} & 
\colhead{$\bF$} & \colhead{$\bar{n}$} & pixels & path length & $E_d$ &
absorption & $\tau_{min}$ & $\tau_{max}$ & lines\\
&&&&&&($\kms$)&&lines&&& used} 
\startdata
3.39 & 4.43 & 3.89 & 0.48 & 0.029 & 35120 & 70893 &8& 281 & 3.8 & 47 &98 \\
2.67 & 3.39 & 2.99 & 0.68 & 0.011 & 35283 & 87308 &12& 284 & 1.0 & 19 &160 \\
2.09 & 2.67 & 2.41 & 0.81 & 0.028 & 36150 & 104581 &9& 223 & 0.41 & 5.4 
& 179 \\
\enddata
\tablecomments{ The minimum (maximum) optical depth of fitted
lines used
in the temperature measurement is given by $\tau_{min}$ 
($\tau_{max}$).
The mean noise level in the spectra is $\bar{n}$.}
\end{deluxetable}
\begin{deluxetable}{lcccc}
\tablecolumns{5}
\tablecaption{Binned Results of Temperature from Observations \label{bigtab}}
\tablehead{
\colhead{$\tau_{c,med}$} & \colhead{$\tau_{c,min}$} & 
\colhead{$B_C$} & \colhead{$\Delta T$} &
\colhead{$\Db$}  \\ 
& &(K)&(K) &} 
\startdata
\cutinhead{$\bar{z}=3.9$}
0.72 & 0.19 & $28200\pm 16700$ & 16900 & 0.48  \\
1.53 & 1.25 & $17400\pm 11200$ & 11100 & 0.67 \\
3.22 & 2.34 & $21200\pm 4100$ & 6100 & 0.97 \\
7.83 & 5.97 & $22400\pm 7900$ & 3100 & 1.66 \\
\cutinhead{$\bar{z}=3.0$}
0.43 & 0.18 & $25800\pm 2200$ & 13800 & 0.61 \\
1.03 & 0.62 & $24600\pm 3700$ & 4400 & 0.90 \\
1.55 & 1.29 & $21800\pm 2100$ & 3000 & 1.15 \\
4.80 & 2.93 & $25600\pm 5600$ & 1600 & 2.53 \\
\cutinhead{$\bar{z}=2.4$}
0.52 & 0.18 & $22600\pm 2600$ & 2300 & 0.99 \\
0.96 & 0.74 & $23200\pm 8500$ & 3000 & 1.44 \\
1.59 & 1.48 & $33000\pm 5400$ & 1000 & 2.07 \\
3.88 & 3.18 & $41600\pm 12100$ & -700 & 4.01 \\
\enddata
\tablecomments{
The temperature cutoff $B_C$, systematic offset $\Delta T$,
and gas density $\Db$,
are given for each bin with minimum optical depth $\tau_{c,min}$ and
median optical depth $\tau_{c,med}$.  }
\end{deluxetable}   
\begin{deluxetable}{lccccccc}
\tablecolumns{9}
\tablecaption{Power-law fits to Temperature from Observations.
\label{plfittab}}
\tablehead{
\colhead{$\bar{z}_{obs}$} & \colhead{$z_{sim}$} & \colhead{$T_\star$}&
\colhead{$T_{1.4}$}  &
\colhead{$\gmo$} & \colhead{$\Delta_\star$} &
\colhead{$T_0$} & \colhead{$\tau_\star$} \\
&&(K)&(K)&&&(K)&}
\startdata
3.89 & 4 & $20200\pm 2700$ &$20100\pm 2800$ & $0.43\pm 0.45$ & $1.42\pm 0.08$ & $17400\pm 3900$ & 6.54 \\
2.99 & 4 & $20200\pm 1300$ &$20300\pm 1400$ & $0.29\pm 0.30$ & $1.37\pm 0.11$ & $18400\pm 2100$ & 1.83 \\
2.99 & 3 & $19600\pm 1500$ &$20400\pm 1800$ & $0.33\pm 0.28$ & $1.24\pm 0.10$ & $18300\pm 1800$ & 1.74 \\
2.41 & 3 & $22600\pm 1900$ &$20700\pm 1900$ & $0.52\pm 0.14$ & $1.66\pm 0.11$ & $17400\pm 1900$ & 1.07 \\
2.41 & 2 & $23400\pm 2000$ &$22800\pm 2100$ & $0.51\pm 0.14$ & $1.47\pm 0.10$ & $19200\pm 2000$ & 0.98 \\
\enddata
\end{deluxetable}

\end{document}